\documentclass{comjnl}

\usepackage[normalem]{ulem}
\usepackage{enumerate}
\usepackage{amsmath}

\usepackage{pdflscape}
\usepackage{afterpage}
\usepackage{capt-of}
\usepackage{graphicx}

\usepackage{amssymb}
\usepackage{graphicx}
\usepackage{subfigure}
\usepackage{multirow}
\usepackage{algorithm}
\usepackage{caption}
\usepackage[noend]{algpseudocode}
\newtheorem{defn}{Definition}
\newtheorem{exmp}{Example}
\usepackage{url}
\begin{document}

\title[A Context-Aware Access Control Policy Framework]{A Policy Model and Framework for Context-Aware Access Control to Information Resources\footnotemark[1]} 

\author{A. S. M. Kayes$^1$, Jun Han$^2$, Wenny Rahayu$^1$, Md. Saiful Islam$^3$, and Alan Colman$^2$}
\affiliation{$^1$La Trobe University, VIC 3086, Australia\\
$^2$Swinburne University of Technology, VIC 3122, Australia\\
$^3$Griffith University, QLD 4215, Australia} 
\email{\{a.kayes, w.rahayu\}@latrobe.edu.au}
\shortauthors{A.S.M. Kayes, J. Han, W. Rahayu, M.S. Islam, and A. Colman}


\begin{abstract}
In today's dynamic ICT environments, the ability to control users' access to information resources and services becomes ever important. On the one hand, it should adapt to the users' changing needs; on the other hand, it should not be compromised. Therefore, it is essential to have a flexible specification of access control polices, incorporating dynamically changing context information. The basic role-based access control (RBAC) approach has been the most widely used access control approach and it typically evaluates access permissions through roles assigned to users who are requesting access to resources. However, it does not provide adequate functionality to incorporate and adapt to context information which could have an impact on access decisions in context-aware environments. Such environments need an access control approach with both dynamic associations of user-role and role-permission capabilities. Towards this end, this paper introduces a policy framework for context-aware access control (CAAC) applications that extends the RBAC approach with context information. The framework uses the relevant context information that reflects the dynamically changing conditions of the environments to specify the CAAC policies: the context-aware user-role and role-permission assignment policies. We first present a {\it formal policy model} for our framework, specifying CAAC policies. Using this model, we then introduce a {\it policy ontology} for modelling CAAC policies and a {\it policy enforcement architecture} which supports access to resources according to the dynamically changing context information. In addition, we {\it evaluate our policy ontology model and framework} by considering (i) the {\it completeness} of the ontology concepts, specifying different context-aware user-role and role-permission assignment policies from the healthcare scenarios; (ii) the {\it correctness} and {\it consistency} of the ontology semantics, assessing the core and domain-specific ontologies through the healthcare case study; and (iii) the {\it performance} of the framework by means of response time. The evaluation results demonstrate the feasibility of our framework and quantify the performance overhead of achieving context-aware access control to information resources. 
\end{abstract}

\maketitle

\section{Introduction}

\footnotetext[1]{An earlier version of this paper appeared under the title ``A Semantic Policy Framework for Context-Aware Access Control Applications" in the {\it Proceedings of 12$^{th}$ IEEE International Conference on Trust, Security and Privacy in Computing and Communications (TrustCom 2013)}. IEEE, pages 753-762, 16-18 July, 2013, Melbourne, Australia \cite{Kayes13}.}

Access control is one of the fundamental security mechanisms needed to protect resources against unauthorized access according to a security policy. It verifies whether a user is allowed to carry out a specific action on a resource. However, the access control decision making processes in today's dynamic environments are not straightforward. In particular, the expected security mechanisms towards this end need to be context-aware in order to cope with highly dynamic environments, thus taking into account the context information \cite{Dey01,DeyAS01} (e.g., the location and request time of the users) so that they can adapt themselves to changing situations. This paradigm shift brings new challenges. On the one hand, users demand access to resources in an anytime, anywhere fashion. On the other hand, such access has to be carefully controlled due to the additional challenges coming with the dynamically changing environments. One such challenge is how to incorporate the dynamically changing context information into the access control policies, to make appropriate and yet possibly different decisions as the user's contextual situation changes.

In the literature, a significant number of access control approaches towards this end have been developed over the last couple of decades ranging from general approaches to application-specific approaches.

A recent study shows that role-based access control (RBAC) \cite{Ferraiolo92, SandhuCFY96, sandhu1994access} has become the most widely used access control approach \cite{Conner}. It has received broad support as a general approach to access control, and is well recognized for its many advantages in large-scale authorization management \cite{Ferraiolo:2001}. RBAC typically evaluates access permission through roles assigned to users and each role assigns a collection of permissions to users who are requesting access to the resources \cite{Ferraiolo92,SandhuCFY96}. It simplifies the management of access control policies by creating user-role and role-permission mappings. However, the basic RBAC approach does not provide adequate functionality to incorporate and adapt to the dynamically changing context information of users. For example, a nurse can access a patient's medical information in the hospital while on duty, but should not access such information while on a public bus heading home.

Context-aware access control is a security mechanism needed to provide flexible control of users' access to resources according to the currently available context information \cite{corradi2004context}. Over the last decade, a number of context-aware access control approaches have been developed, extending the basic RBAC approach by incorporating some specific types of context information: temporal information (e.g., \cite{BertinoBF01}, \cite{JoshiBLG05}), spatial information (e.g., \cite{BertinoCDP05}), and both time and location (e.g., \cite{ChandranJ05}). Recently, several works have extended the basic RBAC approach, considering some further context information other than the {\it temporal} and {\it spatial} dimensions, such as the {\it resource} and {\it environment} dimensions as well as the {\it user} dimension \cite{H11, KulkarniT08, ScheferWenzlS13, HosseinzadehVRL16, TrnkaC16}. These access control approaches are highly domain-specific and still limited in capturing a wide range of important context information for context-aware user-role and role-permission assignments. This gap in the literature suggests that there is a need for a new policy model and framework for {\it context-aware access control} of software services or information resources. 

Consider the roles of {\it patients} and {\it nurses} in a hospital context. A nurse Mary can be allowed to access the medical health records of a patient Bob, if she has been {\it assigned to look after Bob} but only when she is {\it located in the general ward} and {\it during her ward shift time}. That is, Mary can play the {\it nurse} role and consequently can access Bob's {\it medical health records} if these contextual conditions are fulfilled. The {\it contextual conditions} can be used for dynamically performing user-role and role-permission assignments. Therefore, an access controller needs to evaluate such contextual conditions when enabling user-role and role-permission assignments. As such kinds of dynamic attributes (contextual conditions) need to be integrated into the basic role-based access control approach, some important issues arise to realize flexible and dynamic access control. In general, in order to achieve context-awareness and integrate the dynamic attributes into the access control processes, the following research issues need to be addressed: 
\begin{enumerate} [(R1)]
\item How to specify context-aware user-role assignment policies in order to dynamically assign users to roles according to the relevant context information? 
\item How to specify context-aware role-permission assignment policies in order to dynamically assign roles to permissions according to the relevant context information?
\item How to enforce and evaluate context-aware user-role and role-permission assignment policies in order to provide context-aware resource access permissions to users?
\end{enumerate}

\subsection{The Contributions}

To address the above-mentioned research challenges and issues, the overall goal of this paper is to develop a new {\it Context-Aware Access Control (CAAC) Policy Framework}, that is capable of providing secure access to resources and services according to the dynamically changing contexts. It makes the following key contributions:
\begin{enumerate} [(C1)]
\item \textbf{\textit {Formal policy model.}} We propose a {\it formal policy model} that extends the basic RBAC policy model. The novel feature of our policy model is a formal language for specifying the elements of our framework, specifically addressing the following aspects.

\textit{\textbf{(i) Context-aware user-role assignment (CAURA).}} Our policy model uses the relevant {\it context} information to specify context-aware user-role assignment policies. We take {\it context information} to mean any information that can be used to characterize the state of a relevant entity or the state of a relevant relationship between entities.

\textit{\textbf{(ii) Context-aware role-permission assignment (CARPA).}} Our policy model uses the relevant {\it context} information to specify context-aware role-permission assignment policies.
 
\textit{\textbf{(iii) Context specification language (CSL).}} Our policy model includes a simple language ({\it context specification language, $CSL$}) for expressing simple and complex contexts. The simple contexts are the basic low-level context information that is directly obtained from entities and the complex contexts are derived from other basic or complex context information. The simple and complex contexts are used to express the contextual conditions (in the form of contextual expressions) which are required in specifying context-aware user-role and role-permission assignment policies.

\item \textbf{\textit {Policy ontology.}} Based on the above aspects, we introduce a {\it policy ontology} for modelling context-aware access control (CAAC) policies (i.e., user-role and role-permission assignment policies) that take into account the relevant context information. The policy ontology represents the basic elements using the ontology languages OWL and SWRL. 

\item \textbf{\textit {Policy enforcement architecture.}} A policy enforcement architecture (PEA) is introduced that supports context-specific control over access to information resources.

\item \textbf{\textit {Evaluation.}} Other than the above three contributions, this paper also justifies the feasibility of the {\it policy ontology model and framework}. The feasibility is demonstrated by evaluating the {\it completeness} of the ontology model components (ontology concepts), the {\it correctness} and {\it consistency} of the ontology semantics, and the {\it performance} of the policy framework in terms of response time. 

\textbf{\textit {(i) Completeness.}} Covering our policy model features, a detailed case study (including two test cases) from the healthcare domain is presented which demonstrates the completeness of our proposed model components.

\textbf{\textit {(ii) Correctness and Consistency.}} Assessing the base and domain-specific ontologies and executing some queries against the healthcare case study, we have evaluated the correctness and consistency of the ontology model semantics.

\textbf{\textit {(iii) Performance.}} In order to demonstrate the feasibility of our policy framework, we have conducted some sets of experiments in a healthcare environment and quantified the performance overhead of our proposed framework.
\end{enumerate}

The rest of this paper is organized as follows. Section 2 presents an application scenario of a healthcare domain to motivate our work. Section 3 presents a formal model for specifying the elements of our policy framework. In Section 4, we introduce a policy ontology that uses the semantic technologies for specifying context-aware access control policies. In order to support context-specific control over access to information resources, Section 5 introduces a policy enforcement architecture. Section 6 presents the feasibility of our ontology-based policy framework by considering four factors, namely, completeness, correctness, consistency and performance. We review the related work in Section 7. Finally, Section 8 concludes the paper and outlines future work.

\section{Research Motivation}
\label{rm}

In this section, we present a motivating scenario in the domain of patient medical records management and its associated requirements. We consider an extended application scenario from our previous work \cite{KayesHC15CJ}. The objectives of this section are twofold. The first objective is to analyze the scenario which illustrates the need for the incorporation of dynamic contexts in the access control process: context-aware user-role and role-permission assignments. The second objective is to identify the general requirements of developing context-aware access control applications via the context-aware user-role and role-permission assignments. We have used suitable examples from this scenario throughout the paper to explain the concepts of our framework.

\subsection{Motivating Scenario}

Let us consider the area of patient medical records management (PMRM) in the healthcare domain as a motivating scenario (Scene \#1 and Scene \#2) \cite{KayesHC15CJ}.

\begin{itemize}
\item \textit{\textbf{Scene \#1:}} {\it The scenario begins with patient Bob who is in the emergency room due to a heart attack. While not being Bob's usual treating physician, Jane, a general practitioner at the hospital, is required to treat Bob and needs to access Bob's emergency medical records from the emergency room}. 
\end{itemize}

The different types of information involved in this scenario are highly dynamic. Following the traditional RBAC, the access control policy can either be too restrictive and prevent Jane from accessing Bob's emergency medical records (i.e., only the treating physician can), or allow Jane's access but too liberal and potentially compromising privacy (i.e., all general practitioners can). As such, we need context-aware access control (CAAC) policies. One of the relevant policies (in natural language) is shown in Table \ref {table:policy1}.

\begin{table}
\begin{center}
\caption{A CAAC Policy}
\vspace{-0.1in}
\label{table:policy1}
    \begin{tabular}{|p{0.7cm}| p{6.7cm}|}
    \hline
    \bf No & \bf Policy \\ \hline \hline
Policy \#1 & A general practitioner who is a treating doctor of a patient, is allowed to read/write the patient's emergency medical records in the hospital. However, all general practitioners should be able to access the patient's emergency medical records in the hospital (by playing the emergency doctor role), when the patient's health condition is critical. \\ \hline
\end{tabular}
\end{center}
\end{table}

\begin{itemize}
\item \textit{\textbf{Scene \#2:}} {\it After getting emergency treatment, Bob is shifted from the emergency department to the general ward of the hospital and has been assigned a registered nurse, Mary, who has regular follow-up visits to monitor his health condition. Mary needs to access several types of Bob's medical records (daily medical records, past medical history and private medical records) from the general ward with certain conditions}.
\end{itemize}

The different types of dynamically changing information are also involved in this scenario (e.g., the presence of the doctor and the particular nurse-patient relationship). Similar to Scene \#1, the basic RBAC policy model is too restrictive to support access control to Bob's medical records when the context changes (e.g., Bob's health condition is normal). As such, we need context-aware access control (CAAC) policies. The corresponding context-aware access control policies (plain-language policy rules) are shown in Table \ref{table:policy2}.

\begin{table}
\vspace*{-0.1 in}
\begin{center}
\caption{The CAAC Policies }
\vspace{-0.1in}
\label{table:policy2}
    \begin{tabular}{|p{0.7cm}| p{6.7cm}|}
    \hline
    \bf No & \bf Policy \\ \hline \hline
Policy \#2 & A registered nurse within a hospital is granted the right to read/write a patient's daily medical records during her ward duty time and from the location where the patient is located. \\ \hline
Policy \#3 & The nurse is allowed to read the patient's past medical history, if a general practitioner is present at the same location. \\ \hline
Policy \#4 & The nurse can access the patient's private medical records, if she is an assigned nurse of that patient. \\ \hline
\end{tabular}
\end{center}
\vspace{-0.3in}
\end{table}

These context-aware access control policies are based on a set of constraints on the user roles (e.g., general practitioner, emergency doctor, registered nurse) and services/resources (e.g., daily medical records, past medical history, private medical records). These policies need to be evaluated in conjunction with some further dynamic context information (e.g., location, ward duty time, interpersonal relationship).

The above mentioned PMRM (patient medical records management) application is an example of how to realize context-aware access control decisions. In this application scenario, we have only considered some context-aware access control policies for the general practitioners and registered nurses (2 roles). For the overall PMRM application, the number of policies can be up to 500 with respect to 138 different health professionals \cite{ASCO} (i.e., 138 roles). 

\subsection{Scenario Analysis}

In this section, we analyze the application scenario to capture the technical challenges to control access to resources. As different types of context entities and dynamically changing context information are involved in the access control process, a number of important technical challenges arise. The context-specific control over access to patients' medical records on the running scenario works as follows.

Two different types of context information are involved in the application scenario: simple and complex contexts. For example, the {\it identity/role} of the users and the {\it location} of the users are {\it simple contexts}, and the {\it interpersonal relationship} between the user and patient and the {\it health status} of the patient are {\it complex contexts}. The complex contexts are not obtained directly but can be derived from the other available context information. For example, in Scene \#2, the derived relationship or {\it interpersonal relationship} between the registered nurse and patient can be derived from such context information as the {\it nurse's profile}, {\it patient's profile}, etc. In general, {\it how to express the relevant contextual conditions in terms of the different types of dynamic (basic and derived) context information, which are to be integrated into the access control process, is the first technical challenge}.

Normally, only a patient's treating physician is able to access all the patient's electronic health records. In the mentioned emergency scenario (Scene \#1), Jane, while not being the treating doctor, can access Bob's emergency medical records from the emergency room of the hospital by playing the emergency doctor role. That is, Jane can play the emergency doctor role when she is present in the emergency room and Bob's health condition is critical. On the other hand, she also can play the general practitioner role. Thus, rather than having static user-role assignment, {\it how to dynamically assign the roles to users according to relevant contextual conditions is the second technical challenge}.

In general, a registered nurse is granted the right to read a patient's daily medical records within a hospital. But, she should only be able to read the medical records from the location where the patient is located, during her ward duty time. In the application scenario (Scene \#2), Mary, a registered nurse, can access Bob's daily medical records during her ward duty time because she is present with Bob in the general ward. Furthermore, when the situation changes (e.g., Bob's health condition becomes critical again, as in the Scene \#1), decisions on further access requests by Mary to Bob's daily medical records, may change accordingly (e.g., denied). Therefore, {\it how to dynamically assign the permissions to roles based on the relevant contextual conditions is also an important challenge}. 

\subsection{General Requirements}

In this section, we identify the general requirements of developing context-aware access control applications, based on the above motivating scenario. 

\begin{figure} [t]
\centering
\includegraphics[scale=0.37]{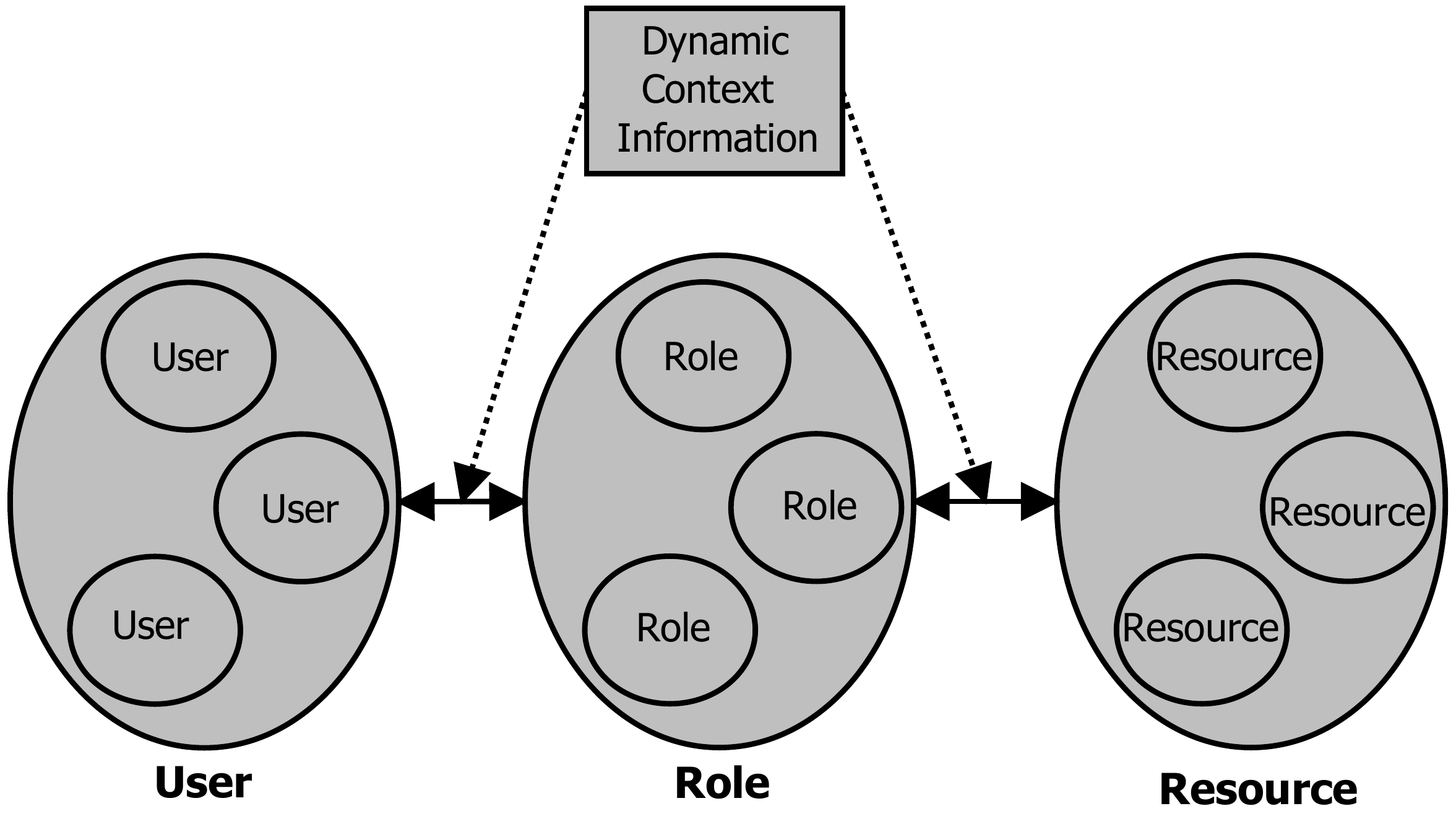}
\caption{The Relationship Chain from User to Resource Access}
\label{fig:genfig}
\end{figure}

Looking at the application scenario and technical challenges identified in the previous sections, we make some important observations concerning context-aware access control. Figure \ref{fig:genfig} illustrates the relationship chain among user, role and resource. 

The relationship chain contains two main parts: the user-role mappings based on the relevant context information, and the role-permission (resource access permission) mappings based on the relevant context information. To support such context-aware access control in a computer application like the medical records management system, we need to consider the 3Ws: \textit{\textbf{who}} (the appropriate users by playing appropriate roles) wants to access \textit{\textbf{what}} (the appropriate parts of the resources), and under \textit{\textbf{what}} contextual conditions (the dynamic context information). In particular, a {\it context-aware access control (CAAC) policy framework} is required to control the access to resources in such applications by taking into account the different types of context information that impact on the access control decisions. The general requirements of developing a CAAC policy framework are as follows:
\begin{enumerate} [(Req.1)]
\item \textit{\textbf{Representation of contextual conditions:}} {\it What access control-specific basic and derived (simple and complex) context information should be considered to express the relevant contextual conditions as part of building context-aware access control?} In general, there is a need to formulate the contextual conditions using the relevant context information, in order to facilitate CAAC use.

\item \textit{\textbf{Specification of user-role assignment policies:}} {\it How to specify the user-role assignment policies based on the relevant contextual conditions?} In general, there are different types of contexts involved in the user-role mappings. For example, in the application scenario, Mary can play the {\it nurse} role under certain contextual conditions (if she is located in the general ward, during her ward shift time and Bob's health condition is normal). If the context/situation changes (e.g., Mary leaves the general ward or Bob's health condition changes from normal to critical), the system will not allow Mary to play the nurse role (relative to Bob). Thus, there is a need to specify dynamic user-role assignment policies based on the relevant contextual conditions.

\item \textit{\textbf{Specification of role-permission assignment policies:}} {\it How to specify the role-permission assignment policies based on the relevant contextual conditions?} For example, in the application scenario, when the context/situation changes (e.g., Bob has come out of emergency and moved to a general ward), decisions on further access control to resources (e.g., Jane's access to Bob's emergency medical records by playing the general practitioner role) may change accordingly (e.g., denied). Thus, there is a need to specify context-aware role-permission assignment policies based on the relevant contextual conditions.

\item \textit{\textbf{Enforcement of access control policies:}} {\it How to evaluate the access control policies based on the relevant contextual conditions to realize a flexible and dynamic access control scheme?} There is a need for an access controller to evaluate the context-aware user-role and role-permission assignment policies based on the relevant contextual conditions, and manage re-authorization of access as context/situation changes.
\end{enumerate} 

\section{Formal Policy Model for CAAC}
\label{fpm}

\subsection{Background}

In this section, we discuss the main elements of the traditional role-based access control model \cite{Ferraiolo92,SandhuCFY96} to motivate the need for extending the basic RBAC model to support context-aware access control requirements.

The traditional RBAC policy model has become the most widely used access control paradigm for managing and enforcing security in large-scale domains \cite{Conner}. The model has the following main elements: users, roles and permissions. In RBAC, the users are human-beings, who are assigned to roles based on their credentials in the organizations, roles represent the job functions within the organizations, describing the authorities and responsibilities conferred on the users assigned to these roles, and  permissions are the approvals to perform certain operations on resources. 

In the RBAC policy model, the access permissions are not assigned directly to the particular users, but to the users' roles. That is, the central notion of RBAC is that users are assigned to appropriate roles, permissions are assigned to roles, and users can have appropriate permissions by being members of such roles. RBAC ensures that only an authorized user is given access to a certain permission, and is based on the user's role in the organization. As such, it simplifies the management of access control policies by creating user-role and role-permission assignments. However, the RBAC policy model does not directly adapt to the requirements of the context-aware access control applications (as presented in the previous section). In particular, it does not provide adequate functionalities to incorporate and adapt to the dynamically changing context information. 

In the next section, we introduce a new context-aware access control (CAAC) policy framework for information resources and software services. It extends  the basic concepts of RBAC to use dynamic context information while making access control decisions. The basic concepts of our policy framework are that users are {\it dynamically} assigned to roles by satisfying the relevant contextual conditions, permissions (resource/service access permissions) are {\it dynamically} assigned to roles by satisfying the relevant contextual conditions, and users acquire appropriate permissions by being members of such roles in a dynamic manner.

\subsection{Core Policy Model}

\begin{figure}[t]
\centering
\includegraphics[scale=0.4]{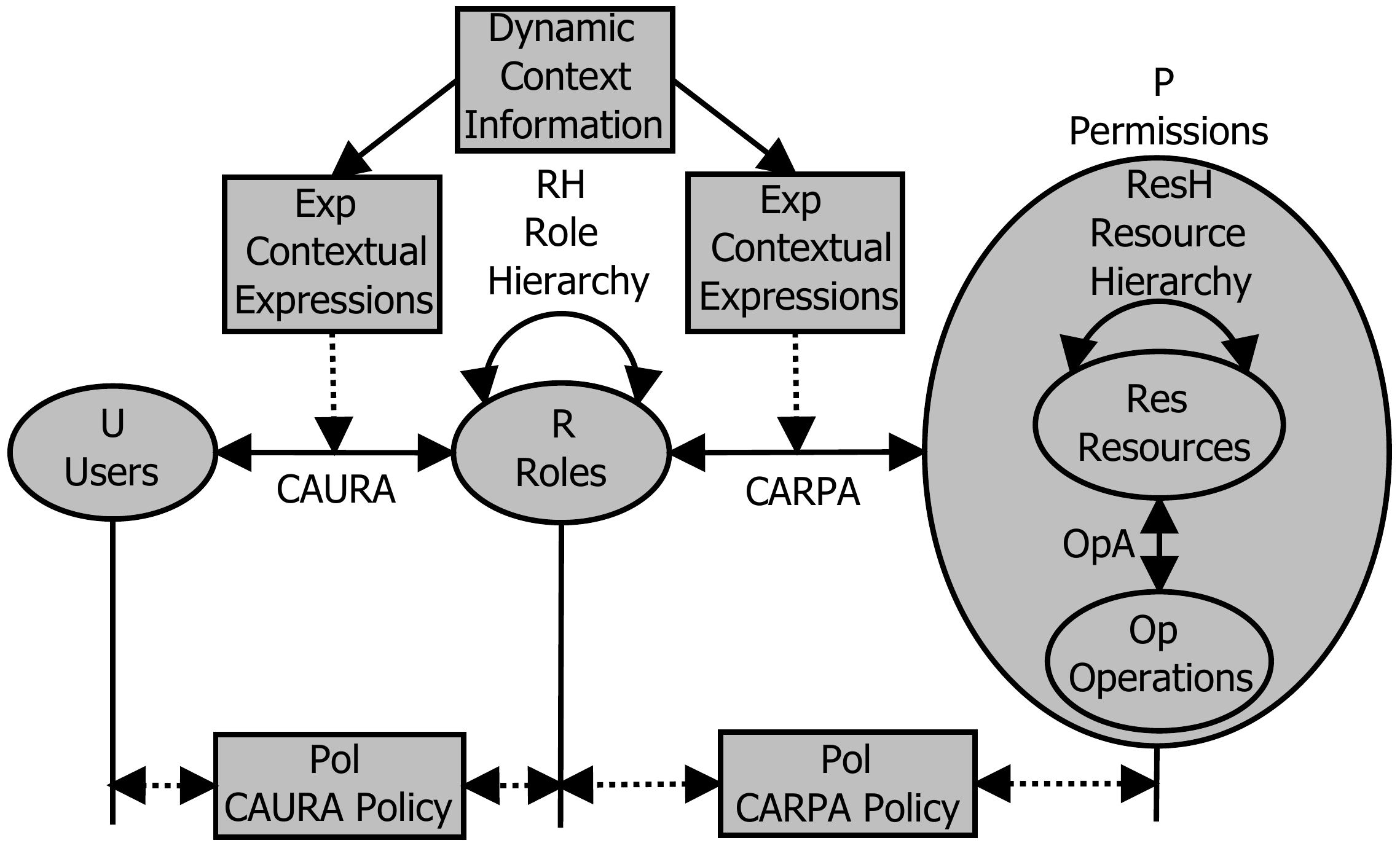}
\caption{Core Policy Model}
\label{fig1:caac}
\end{figure}

Figure \ref{fig1:caac} shows the policy model of our CAAC framework and the relationships between its elements. 

Our policy model enables dynamic privileges assignment in two steps, letting users to access resources and services when certain contextual conditions are satisfied. In the first step, the users are dynamically assigned to the roles when a set of conditions are satisfied. In the next step, when the roles are activated, the permissions (service access permissions) are assigned to these roles when specific contextual conditions are satisfied. The two main concepts are {\it context-aware user-role assignments} and {\it context-aware role-permission assignments}.

Based on the formalization of the traditional role-based access control (RBAC) model \cite{SandhuCFY96}, we present a formal definition of our policy model.

\begin{defn}
(Core Policy Model). The CAAC policy model has the following elements:
\begin{itemize}
\item Basic elements: $U, R, Res, Op$, and $Exp$, \\for users, roles, resources, operations and contextual condition expressions, respectively;
\item Composite elements: \\$RH, ResH, OpA, P, CAURA, CARPA$, and $Pol$, \\for role hierarchy, resource hierarchy, operation assignments, permissions, context-aware user-role assignments, context-aware role-permission assignments and policies, respectively.
\end{itemize}
These elements are explained and further defined below.
\end{defn}

First of all, we define the five basic elements of our policy model:

\begin{itemize}
\item \textit{\textbf{Users (U):}} {\it U} represents a set of users. The users are service requesters interacting with a computing system, whose access requests are being controlled.
\begin{equation}
\begin{split}
U = \{u_{1}, u_{2}, u_{3}, ..., u_{m}\}
\end{split}
\end{equation}

\item \textit{\textbf{Roles (R):}} {\it R} represents a set of roles. The roles reflect users' job functions within the organization (e.g., healthcare domain). 
\begin{equation}
\begin{split}
R = \{r_{1}, r_{2}, r_{3}, ..., r_{n}\}
\end{split}
\end{equation}

\item \textit{\textbf{Resources (Res):}} {\it Res} represents a set of resources. The resources are the objects protected by access control that represent the data/information container (e.g., the different parts of a patient's medical records).
\begin{equation}
\begin{split}
Res = \{res_{1}, res_{2}, res_{3}, ..., res_{o}\}
\end{split}
\end{equation}

\item \textit{\textbf{Operations (Op):}} {\it Op} represents a set of operations on the resources. The operations are the actions that can be executed on the resources, for instance, read and write.
\begin{equation}
\begin{split}
Op = \{op_{1}, op_{2}, op_{3}, ..., op_{p}\}
\end{split}
\end{equation}

\item \textit{\textbf{Expressions (Exp):}} {\it Exp} represents a set of contextual expressions. The expressions are used to express the contextual conditions (using relevant context information) in order to specify the context-aware user-role and role-permission assignment policies. We use the terms of contextual expression and contextual condition interchangeably. The contextual expressions are specified in the context specification language (CSL).
\begin{equation}
\begin{split}
Exp = \{exp_{1}, exp_{2}, exp_{3}, ..., exp_{q}\}
\end{split}
\end{equation}
\end{itemize}

A detailed analysis of contextual expressions is given in the next section (see Section \ref{csl}). \\

The policy model has seven other elements (which are related to the above basic elements). These elements are defined formally as follows:\\

\begin{itemize}
\item \textit{\textbf{Role Hierarchy (RH):}} {\it RH} is a partial order on {\it R} to serve as the role hierarchy, which supports the concept of role inheritance. The role is considered in a hierarchical manner in that if a permission assigned to a junior role, then it is also assigned to all the senior roles of that role.
\begin{equation}
\begin{split}
RH \subseteq R \times R
\end{split}
\end{equation}

\item \textit{\textbf{Resource Hierarchy (ResH):}} {\it ResH} is a partial order on $Res$ to serve as the resource hierarchy, which supports a user's access to the resources at different granularity levels. The resource is considered in a hierarchical manner in that if a user has the right to access a resource at a higher granularity level, then he also has the right to access that part of resource at a lower granularity levels. 
\begin{equation}
\begin{split}
ResH \subseteq Res \times Res
\end{split}
\end{equation}

\item \textit{\textbf{Operation Assignment (OpA):}} {\it OpA} is a many-to-many operation-to-resource mapping. Each operation could be associated with many resources, and each resource could be manipulated by many operations.
\begin{equation}
\begin{split}
OpA \subseteq Res \times Op
\end{split}
\end{equation}

\item \textit{\textbf{Permission (P):}} {\it P} represents a set of permissions. The permissions are the approvals to perform certain operations on resources by the users who initiate access requests.
\begin{equation}
\begin{split}
P = \{(res_{i}, op_{j}) | res_{i} \in Res, op_{j} \in Op\}
\end{split}
\end{equation}

Where i = \{1, 2, 3, ..., o\}, j = \{1, 2, 3, ..., p\}, {\it Res} is a set of resources, and {\it Op} is a set of operations on the resources.\\

The permission ({\it P}) is a subset of operation assignment ({\it OpA}),
\begin{equation}
\begin{split}
P \subseteq OpA
\end{split}
\end{equation}

\item \textit{\textbf{Context-Aware User-Role Assignment (CAURA):}} {\it CAURA} is a context-aware user-role assignment relation, which is a many-to-many mapping between users and roles, when a set of dynamic contextual conditions are satisfied.
\begin{equation}
\begin{split}
CAURA \subseteq U  \times R  \times Exp
\end{split}
\end{equation}

\item \textit{\textbf{Context-Aware Role-Permission Assignment (CARPA):}} {\it CARPA} is a context-aware role-permission assignment relation, which is a many-to-many mapping between roles and permissions, when a set of dynamic contextual conditions (contextual expressions) are satisfied.
\begin{equation}
\begin{split}
CARPA \subseteq R \times P \times Exp
\end{split}
\end{equation}

\item \textit{\textbf{Policies (Pol):}} {\it Pol} represents a set of context-aware access control policies. It includes the context-aware user-role assignment (CAURA) policies and context-aware role-permission assignment (CARPA) policies.
\begin{equation}
\begin{split}
Pol = Pol_{CAURA} \cup Pol_{CARPA}
\end{split}
\end{equation}
\end{itemize}

The main concepts that we have introduced in our policy model are the {\it context-aware user-role assignments} and {\it context-aware role-permission assignments}. They incorporate dynamically changing contextual conditions (in the form of contextual expressions) into RBAC for dynamic user-role and role-permission assignments. 

In the following sections, we further define and discuss {\it context information}, {\it context specification language}, {\it CAURA policy specification} and {\it CARPA policy specification}.

\subsection{Context Information and Context Specification Language (CSL)}
\label{csl}

\subsubsection{Context Information}

In the context-awareness literature, many researchers have defined the concept of context. According to Dey, context is any information that can be used to characterize the situation of an entity (person, place or object) \cite{Dey01}. For our purpose, however, existing context definitions are not specific enough to specify the different types of entities for access control and the context information characterizing these entities. In our earlier work \cite{KayesOnt13, KayesHC15CJ}, we define context as {\it any relevant information about the state of an entity relevant to access control or the state of a relevant relationships between persons (as entities)}.

We classify context information into {\it simple context} and {\it complex context}, i.e., context information (C) is the set of all simple contexts ($C_{s}$) and all complex contexts ($C_{c}$).
\begin{equation}
\begin{split}
C\hspace*{0.05 in}= \hspace*{0.05 in} C_{s} \hspace*{0.05 in} \cup \hspace*{0.05 in}C_{c}
\end{split}
\end{equation}

A simple context `$c_{s}$' ($c_{s} \in C_{s}$) is an attribute of an entity that directly depends on a raw context fact. It characterizes the state of an entity, based on a single context information source.
\begin{equation}
\begin{split}
C_{s} = \{c_{s1}, c_{s2}, c_{s3}, ..., c_{ss}\}
\end{split}
\end{equation}

In our application scenario (presented in Section \ref{rm}), {\it user identity} is a simple context that represents a property of the user (resource requester).

A complex context `$c_{c}$' ($c_{c} \in C_{c}$) is a combination of context facts. It depends on the values of attributes that characterize the state of one or more entities, based on the one or more context information sources.
\begin{equation}
\begin{split}
C_{c} = \{c_{c1}, c_{c2}, c_{c3}, ..., c_{ct}\}
\end{split}
\end{equation}

In our application scenario, {\it interpersonal relationship} is a complex context that represents a property which is related to the user and resource owner. The {\it interpersonal relationship} between user and owner can be inferred from available context information (e.g., the profile information of the user and owner).

\subsubsection{Context Specification Language}

Our policy model includes a simple language (context specification language, $CSL$) for expressing contextual conditions based on the simple and complex contexts. This language is used to formally specify the constraints in context-aware user-role and role-permission assignment policies.

\begin{defn}
\label{sc}
(Simple Context Expression ($Exp_{s}$)). Let $E$ be the set of context entities, and $C_{s}$ be the set of simple context information, then we define a simple context expression as a tuple in the form of
\begin{equation}
\begin{split}
<e.c_{s}, rel.op, v>, [ where, e \in E, c_{s} \in C_{s}, and \\rel.op \in \{<, \leq, >, \geq, =, \neq\}]
\end{split}
\end{equation}
\end{defn}

In the above expression, `$e$' denotes a context entity, `$c_{s}$' denotes a simple context attribute, `$e.c_{s}$' denotes a simple context about an entity (i.e., context attribute of an entity), `$rel.op$' denotes a relational operator (the set of `$rel.op$' can be extended to accommodate user-defined operators (e.g., `$entering$'), and `$v$' denotes the value assigned to the context attribute `$c_{s}$' of context entity `$e$'. 

\begin{exmp}
A patient or resource owner's heart rate is less than 65 (or is abnormal), which is represented as,
\begin{equation}
\begin{split}
exp_{s1} \triangleq (Owner.heartRate < 65) ~or \\exp_{s1} \triangleq (Owner.heartRate = ``Abnormal")
\end{split}
\end{equation}
\end{exmp}

A simple context expression is a simple contextual constraint in the {\it CSL} language. It is possible to construct more complex expressions by logically combining (conjunction ($\wedge$), disjunction ($\vee$), and negation ($\neg$)) simple constraints.

\begin{defn}
\label{cc}
(Complex Context Expression ($Exp_{c}$)). A complex context expression can be defined by performing logical composition on simple or complex context expressions.
\begin{equation}
\begin{split}
exp_{c1}\hspace*{0.1 in} \triangleq \hspace*{0.03 in}exp_{1} \hspace*{0.05 in} \wedge \hspace*{0.05 in} exp_{2}\\
exp_{c2}\hspace*{0.1 in} \triangleq \hspace*{0.03 in}exp_{3} \hspace*{0.05 in} \vee \hspace*{0.05 in} exp_{4}\\
exp_{c3}\hspace*{0.1 in} \triangleq \hspace*{0.03 in}\neg \hspace*{0.05 in}exp_{5}
\end{split}
\end{equation}
\end{defn}

where $exp_{1}$, $exp_{2}$, $exp_{3}$, $exp_{4}$, and $exp_{5}$ are already defined simple or complex context expressions, and $exp_{c1}$, $exp_{c2}$, and $exp_{c3}$ are newly defined complex context expressions.

\begin{exmp}
Let us consider an access control policy from our application scenario: Mary can play the registered nurse role during her ward duty time and when she is located in the general ward. Using logical composition, the contextual condition of this policy can be represented as 
\begin{equation}
\label{rncc1}
\begin{split}
exp_{c1} \triangleq ((User.locationAddress = ``GeneralWard") \\\wedge~ (User.requestTime = ``DutyTime"))
\end{split}
\end{equation}
\end{exmp} 

\begin{exmp}
A registered nurse (who is assigned for a patient) is granted the right to access the patient's daily medical records when the patient's health condition is normal. Using logical composition, the contextual condition of this policy can be represented as 
\begin{equation}
\label{rncc2}
\begin{split}
exp_{c2} \triangleq ((interRelationship(User, Owner) = \\``AssignedNurse") \wedge (Owner.healthStatus = \\``Normal"))
\end{split}
\end{equation}
\end{exmp} 

\begin{defn}
(Contextual Expression ($Exp$)). By definitions \ref{sc} and \ref{cc}, a contextual expression $exp$ ($exp \in Exp$) is either a simple context expression or a complex context expression.
\begin{equation}
\begin{split}
\label{exp}
exp \hspace*{0.1 in} \triangleq \hspace*{0.1 in} exp_{s} \hspace*{0.1 in} | \hspace*{0.1 in} exp_{c}
\end{split}
\end{equation} 
\end{defn}

Where $exp_{s}$ denotes a simple context expression and $exp_{c}$ denotes a complex context expression.

\subsection{Context-Aware User-Role Assignment (CAURA) Policy Specification}
\label{caurap}

Our policy model extends the concept of user-role assignment in RBAC, by introducing the concept of context-aware user-role assignment ($CAURA$). The traditional RBAC model defines user-role assignment ($URA$) simply as a mapping of users to roles. 
\begin{align}
URA \subseteq U \times R
\end{align}
We have extended this $URA$ notion by introducing dynamic contextual expressions (integrating relevant context information).

\begin{defn}
(CAURA). Let $U$ be the set of users, $R$ be the set of roles and $Exp$ be the set of contextual expressions, then CAURA is a many-to-many user-role assignment relation associated with certain contextual expressions. 
\begin{equation}
\begin{split}
CAURA = \{(u_{1}, r_{1}, exp_{1}), (u_{2}, r_{2}, exp_{2}), \\...~,~ (u_{i}, r_{j}, exp_{k})\} \subseteq U \times R \times Exp
\end{split}
\end{equation}
\end{defn}
where `$u$' denotes a user ($u \in U$), `$r$' denotes a role ($r \in R$) and `$exp$' denotes a contextual expression ($exp \in Exp$).

Context-aware user-role assignments can be expressed in tabular form (see Table \ref{t:rr}). For example, the second row in Table \ref{t:rr} describes when Mary is present in the general ward and during her ward duty time, she can be assigned to the registered nurse role.

\begin{table}
\begin{center}
\caption{Context-Aware User-Role Assignment}
\label{t:rr}
 \begin{tabular}{|p{0.8cm} |p{2.6cm} |p{4cm}|}
    \hline
    \textbf{\textit {User}} & \textbf{\textit {Role}} & \textbf{\textit {Contextual Expression}} \\ \hline
   $ Jane$ & $EmergencyDoctor$ & $(User.locationAddress = ``EmergencyRoom")$ \\  \hline
    $Mary$ & $RegisteredNurse$ & $(User.locationAddress = ``GeneralWard")$ $\wedge$ $(User.requestTime = ``DutyTime")$ \\
 \hline
\end{tabular}
\end{center}
\vspace{-0.2in}
\end{table}

\begin{defn}
\label{caua}
(CAURA Policy). Let $U$ be the set of users, $R$ be the set of roles and $Exp$ be the set of contextual expressions. A $CAURA$ policy ($pol_{CAURA}$) is defined as follows: 

\begin{quote}
{\it a user `$u$' can be assigned the role `$r$' \\if and only if $(u, r, exp) \in CAURA$ \\or alternatively\\
$pol_{CAURA}$ $\in$ $CAURAPolicy$ \\$\iff$ $(u, r, exp)$ $\in$ $CAURA$}
\end{quote}
where `$u$' is a user ($u \in U$), `$r$' is a role ($r \in R$), and `$exp$' is a contextual expression ($exp \in Exp$) defined in the $CSL$.
\end{defn}

\begin{table*}
\begin{center}
\caption{Context-Aware User-Role Assignment Policy}
\label{table:caua}
    \begin{tabular}{|p{13cm}|}
    \hline
\hspace*{0.05 in}$\textbf{\textit {If}}$\\
\hspace*{0.3 in}$CAURAPolicy(caura)$\hspace*{0.05 in}$\wedge$\hspace*{0.05 in}$User(u)$\hspace*{0.05 in}$\wedge$\hspace*{0.05 in}
$Role(r)$\hspace*{0.05 in}$\wedge$ 
\hspace*{0.05 in}$ContextualCondition(exp)$ $\wedge$ \\
\hspace*{0.3 in}$hasUser(caura, u)$ \hspace*{0.05in}$\wedge$\hspace*{0.05 in}$hasRole(caura, r)$
\hspace*{0.05 in}$\wedge$ \hspace*{0.05 in}$hasCondition(caura, exp)$
\hspace*{0.05 in}\\
\hspace*{0.05 in}$\textbf{\textit {Then}}$\\
\hspace*{0.3 in}$caura(u, r)$\\
    \hline
\end{tabular}
\end{center}
\end{table*}

\begin{table*}
\begin{center}
\caption{Context-Aware Role-Permission Assignment Policy}
\label{table:capa}
    \begin{tabular}{|p{13cm}|}
    \hline
\hspace*{0.05 in}$\textbf{\textit {If}}$\\
\hspace*{0.3 in}$CARPAPolicy(carpa)$\hspace*{0.05 in}$\wedge$\hspace*{0.05 in}$Role(r)$\hspace*{0.05 in}$\wedge$\hspace*{0.05 in}
$Resource(res)$\hspace*{0.05 in}$\wedge$
\hspace*{0.05 in}$Operation(op)$\hspace*{0.05 in}$\wedge$ \\ 
\hspace*{0.3 in}$ContextualCondition(exp)$ \hspace*{0.05 in}$\wedge$ $hasRole(carpa, r)$\hspace*{0.05 in}$\wedge$
\hspace*{0.05 in}$hasResource(carpa, res)$\hspace*{0.05 in}$\wedge$\\
\hspace*{0.3 in}$hasOperation(carpa, op)$\hspace*{0.05 in}$\wedge$
\hspace*{0.05 in}$hasCondition(carpa, exp)$\hspace*{0.05 in}\\
\hspace*{0.05 in}$\textbf{\textit {Then}}$\\
\hspace*{0.3 in}$carpa(r, p)$\\
    \hline
\end{tabular}
\end{center}
\vspace{-0.1in}
\end{table*}

Based on the CAURA policy definition (see Definition \ref{caua}), the rule (shown in Table \ref{table:caua}) expresses the context-aware user-role assignment policy, i.e., a User $u$ ($u \in U$) can play a Role $r$ ($r \in R$) under contextual condition $exp$ ($exp \in Exp$).

\begin{exmp}
Let us consider an access control policy from our application scenario: {\it Mary can play the registered nurse role during her ward shift time and when she is located in the ward}. In Table \ref{table:caua}, Role `$r$' denotes $RegisteredNurse$ role, User `$u$' is $Mary$, and Contextual Condition `$exp$' is the combination of the raw facts of the ward `duty time' and `location' of Mary. The contextual condition is already expressed in Formula (\ref{rncc1}) in CSL (Section \ref{csl}).
\end{exmp}

\subsection{Context-Aware Role-Permission Assignment (CARPA) Policy Specification}
\label{carpap}

Similar to context-aware user-role assignment, our policy model extends the concept of role-permission assignment in RBAC with contextual expressions, called context-aware role-permission assignment ($CARPA$).
Traditional RBAC model defines role-permission assignment ($RPA$) simply as a mapping of roles to permissions. 
\begin{align}
RPA \subseteq R \times P
\end{align}
We have extended this $RPA$ notion by introducing dynamic contextual expressions (integrating context information).

\begin{defn}
(CARPA). Let $R$ be the set of roles, $Exp$ be the set of contextual expressions, and $P$ be the set of permissions, then CARPA is a many-to-many role-permission assignment relation associated with certain contextual expressions.
\begin{equation}
\begin{split}
CARPA = \{(r_{1}, p_{1}, exp_{1}), (r_{2}, p_{2}, exp_{2}), \\...~, ~(r_{i}, p_{j}, exp_{k})\} \subseteq R \times P \times Exp
\end{split}
\end{equation}
\end{defn}
where `$p$' denotes a permission ($p \in P$), `$r$' denotes a role ($r \in R$) and `$exp$' denotes a contextual expression ($exp \in Exp$).
\\
\begin{defn}
(CARPA Policy). Let $R$ be the set of roles, $P$ be the set of permissions, and $Exp$ be the set of contextual expressions. A $CARPA$ policy ($pol_{CARPA}$) is defined as follows:

\begin{quote}
{\it a role `$r$' can be assigned the permission `$p$' \\if and only if $(r, p, exp) \in CARPA$ \\or alternatively \\
$pol_{CARPA}$ $\in$ $CARPAPolicy$ \\$\iff$  $(r, p, exp)$ $\in$ $CARPA$}
\end{quote}
where `$r$' is a role ($r \in R$), `$p$' is a permission ($p \in P$), and `$exp$' is a contextual expression ($exp \in Exp$) defined in the $CSL$.
\end{defn}

Based on the CARPA policy definition, the following rule (shown in Table \ref{table:capa}) expresses the context-aware role-permission assignment policy.

\begin{exmp}
In our application scenario, let us consider an access control policy: {\it A registered nurse, who is assigned for a regular follow-up visit to monitor a patient's health condition, can access the patient's daily medical records (DMR) when the patient's health status is normal}. In Table \ref{table:capa}, Role `$r$' denotes $RegisteredNurse$ role, permission is {\it (DMR, Write)} (i.e., Resource `$res$' is $DMR$ and Operation `$op$' is $Write$), and Contextual Condition `$exp$' is the combination of the following raw facts: the nurse is `assigned' for the patient and the patient has `normal health status'. The contextual condition is already expressed in Formula (\ref{rncc2}) in the CSL (Section \ref{csl}).
\end{exmp}

\section{Ontology-Based Policy Model for Policy Specification}
\label{po}

We have in the last section defined the formal policy model to specify context-aware user-role and role-permission assignment policies. Based on this formal model, in this section, we present an ontology-based policy model for our framework to provide the practical basis for realizing our CAAC framework, as the policy ontology can be directly included in its implementation (see the next section). The main goal of our policy ontology is to specify the two sets of context-aware access control policies by incorporating the dynamically changing context information. 

\subsection{Design Considerations}

\begin{table*} 
\begin{center}
\caption{Domain and Range Restrictions for CAURA Object Properties}
\label{drcaura}
    \begin{tabular}{|p{3cm}|p{3cm}|p{3cm}|p{6.8cm}|}
    \hline
    \bf Object Property & \bf Domain & \bf Range & \bf Description \\ \hline \hline
hasCondition & CAURAPolicy & ContextualCondition & A CAURA policy has the contextual conditions \\ \hline
hasRole & CAURAPolicy & Role & A CAURA policy has the roles \\ \hline
hasUser & CAURAPolicy & User & A CAURA policy has the users \\ \hline
hasContext & ContextualCondition & ContextInfo & A contextual condition is formed by the context information which can be either a simple context or a complex context \\ \hline
plays & User & Role & A user can play a role \\ \hline
\end{tabular}
\end{center}
\end{table*}

To simplify the management of access control policies, various policy languages have been proposed in the literature. Our goal in this paper is to provide a way in which context-aware access control policies can be specified, which incorporate context information. To be of practical use, it must be expressive enough to specify the policies in an easy and natural manner. Experience from existing research (e.g., \cite{BettiniBHINRR10}, \cite{RiboniB11}) shows that ontologies are very suitable for modeling dynamic information for ubiquitous computing applications. Furthermore, the expressivity of the ontology language OWL \cite{owl} can be extended by incorporating SWRL rules \cite{swrl}. As such, we use the ontology languages OWL and SWRL as the CAAC policy language.

Our policy ontology specifies two sets of context-aware access control (CAAC) policies: (i) {\it the context-aware user-role assignment policies} and (ii) {\it the context-aware role-permission assignment policies}. The basic concepts of these CAAC policies are that {\it users} are dynamically assigned to {\it roles} by satisfying the {\it relevant contextual conditions}, {\it permissions} are dynamically assigned to {\it roles} by satisfying the {\it relevant contextual conditions}, and users acquire resource access permissions by having corresponding roles. 

\begin{figure}[h!]
\centering
\includegraphics[scale=1]{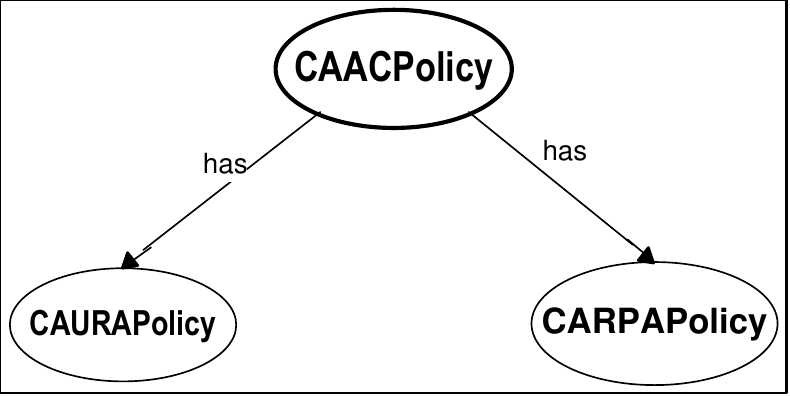}
\caption{The Core Policy Ontology}
\label{f:pocaac}
\end{figure}

Figure \ref{f:pocaac} shows the top-level conceptual view of our policy ontology. The ontology defines the following concepts under the hierarchy of {\it CAACPolicy}, namely {\it CAURAPolicy} and {\it CARPAPolicy}. The {\it CAURAPolicy} models context-aware user-role assignment policies and {\it CARPAPolicy} models context-aware role-permission assignment policies.

\subsection{CAURA Policy Ontology}
\label{caurapo}

The CAURA policy ontology, representing {\it context-aware user-role assignment policies}, has been designed by answering the following questions.
\begin{itemize}
\item Who is requesting resource/service access (requester or user)?
\item What role does the user play (role)?
\item What is the dynamic context information that is relevant for this user-role assignment (contextual condition)? 
\end{itemize}

\begin{figure}[h!]
\centering
\includegraphics [scale=0.8] {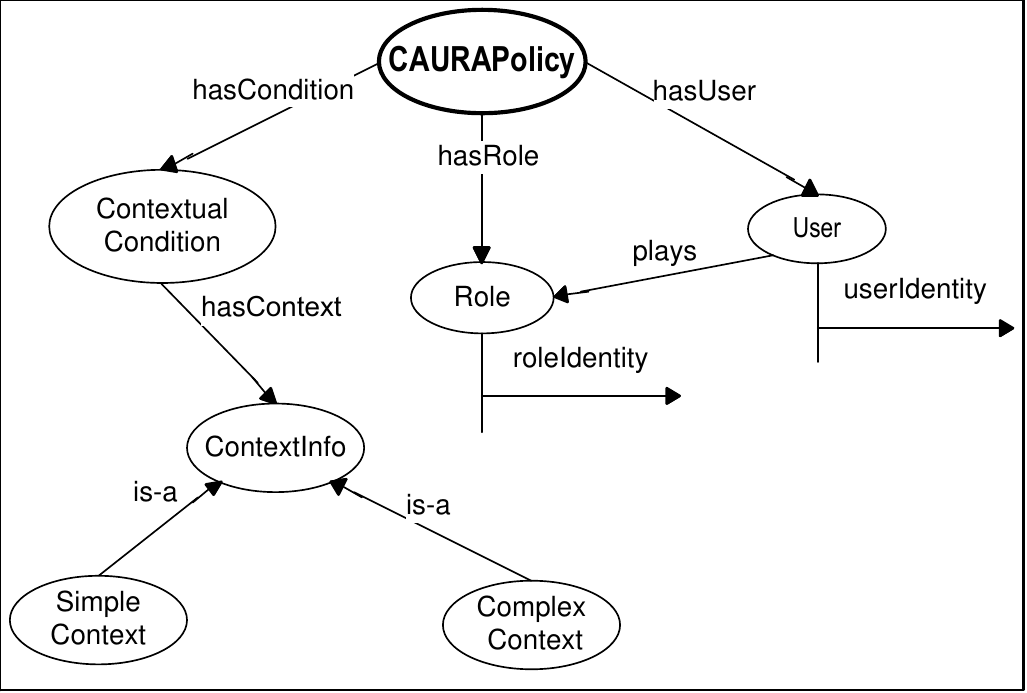}
\caption{The CAURA Policy Ontology}
\label{caurao}
\end{figure}

\subsubsection{Core Concepts}

The {\it CAURA} policy ontology, as depicted in Figure \ref{caurao}, has the following core concepts which are organized into a {\it CAURAPolicy} hierarchy: {\it User}, {\it Role} and {\it ContextualCondition}. The {\it ContextualCondition} class has the concept {\it ContextInfo}, which has two subclasses {\it SimpleContext} and {\it ComplexContext}.

For simplicity, we define the CAURA ontology concepts as follows:

The basic classes {\it User}, {\it Role} and {\it ContextualCondition} form a {\it CAURAPolicy} (see Definition \ref{settt1}).

\begin{quote}
\vspace{-0.2in}
\begin{defn}
(Basic Class).\\
\label{settt1}
\hspace*{0.4 in} $CAURAPolicy \subseteq User  \times Role  \times \\
\hspace*{1.6 in} ContextualCondition$
\end{defn}
\end{quote}

The {\it ContextInfo} class has two subclasses {\it SimpleContext} and {\it ComplexContext} (see Definition \ref{settt3}). A subclass {\it SimpleContext} or {\it ComplexContext} can have fewer or equal elements to the basic class {\it ContextInfo}.

\begin{quote}
\vspace{-0.2in}
\begin{defn}
(Subclass).\\
\label{settt3}
\hspace*{0.4 in}${\it SimpleContext} \subseteq {\it ContextInfo}$\\
\hspace*{0.4 in}${\it ComplexContext} \subseteq {\it ContextInfo}$
\end{defn} 
\end{quote}

The relationships in {\it CAURAPolicy} ontology are represented by two types of properties, i.e., {\it object} and {\it data type} properties. 

The domain and range of object properties are specified in Table \ref{drcaura}.

The class {\it User} has {\it userIdentity} data property and the {\it Role} class has a data property named {\it roleIdentity} (see Table \ref{dtpcaura}).

\begin{table}
\begin{center}
\caption{CAURA Data Type Properties}
\label{dtpcaura}
    \begin{tabular}{|p{3.8cm}|p{3.5cm}|}
    \hline
    \bf Data Type Property & \bf Domain \\ \hline \hline
userIdentity & User\\ \hline
roleIdentity & Role \\ \hline
\end{tabular}
\end{center}
\end{table}

Overall, a CAURA policy captures the \textit{\textbf{3W}} dimensions which can be read as follows: 

\vspace*{0.1 in}
\begin{quote}
{\it a {\it CAURAPolicy} specifies that a {\it user} ({\it userIdentity}) can play a {\it role} ({\it roleIdentity}) by satisfying the relevant {\it contextual conditions.}} 
\end{quote}

The details of OWL-based CAURA policy specifications can be found in Appendix A.

\subsubsection{An Example CAURA Policy}
\vspace*{0.2 in}

\begin{exmp}
\label{rnurp}
Let us consider an access control policy for the registered nurse presented in Section \ref{rm}: a user Mary can play the registered nurse (RN) role (in order to access a patient's daily medical records), during her ward duty time (DT) from the general ward (GW) of the hospital, where the patient is located.
\end{exmp}

\begin{table*}
\begin{center}
\caption{An Example CAURA Policy for Playing Registered Nurse Role}
\vspace{-0.1in}
\label{table:rnurp}
    \begin{tabular}{|p{0.7cm}|p{12cm}|}
    \hline
1 &\hspace*{0.1 in}$<${\bf CAURAPolicy} rdf:ID=``$caura_{1}$"$>$\\
2&\hspace*{0.4 in}$<$hasUser rdf:resource=``\#User\_RegisteredNurse\_DB"/$>$\\
3&\hspace*{0.4 in}$<$hasRole rdf:resource=``\#Role\_RegisteredNurse"/$>$\\
4&\hspace*{0.4 in}$<$hasCondition rdf:resource=``\#ContextualCondition\_ContextInfo"/$>$\\
6&\hspace*{0.1 in}$<${\bf /CAURAPolicy}$>$\\ \hline
7&\hspace*{0.1 in}$<${\bf User} rdf:ID=``User\_RegisteredNurse\_DB"$>$\\
8&\hspace*{0.4 in}$<$userIdentity rdf:datatype=``\&xsd;string"$>$Mary00X$<$/userIdentity$>$\\
9&\hspace*{0.1 in}$<${\bf /User}$>$\\ \hline
10&\hspace*{0.1 in}$<${\bf Role} rdf:ID=``Role\_RegisteredNurse"$>$\\
11&\hspace*{0.4 in}$<$roleIdentity rdf:datatype=``\&xsd;string"$>$RN00X$<$/roleIdentity$>$\\
12&\hspace*{0.1 in}$<${\bf /Role}$>$\\ \hline
13&\hspace*{0.1 in}$<${\bf ContextualCondition} rdf:ID=``ContextualCondition\_ContextInfo"$>$\\
14&\hspace*{0.4 in}$<$hasContext rdf:resource=``\#ComplexContext\_$c_{c1}$"/$>$\\ 
15&\hspace*{0.1 in}$<${\bf /ContextualCondition}$>$
\\ \hline
\end{tabular}
\end{center}
\end{table*}

In this policy, the access decision is based on the following policy constraints: \textit{\textbf{who}} the user is (user's {\it identity}), \textit{\textbf{what}} role she can play (role's {\it identity}), and \textit{\textbf{under what contextual conditions}} (the {\it locations} of the user and patient, and the {\it request time} of the user). The CAURA policy for the registered nurses in OWL is shown in Table \ref{table:rnurp}. The core policy concepts are specified in {\it Line\# 1 to 6}, the user specification is shown in {\it Line\# 7 to 9}, the role specification (registered nurse) is shown in {\it Line\# 10 to 12}, and the contextual condition (a complex context $c_{c1}$) is specified in {\it Line\# 13 to 15}. 

The contextual condition $c_{c1}$ {\it (registered nurses during ward duty time from the general ward of the hospital}) is already expressed in Formula (\ref{rncc1}), using logical composition (see Section \ref{csl}). The following OWL code shows its ontological definition (see Definition \ref{cc1owl}).

\begin{defn}
(`$c_{c1}$' Contextual Condition Definition).
\label{cc1owl}
\end{defn}
{\it \hspace*{0.2 in}$<$ComplexContext rdf:ID=``ComplexContext\_$c_{c1}$"$>$\\
\hspace*{0.6 in}$<$User.locationAddress rdf:datatype\\\hspace*{1 in}=``\&xsd;string"$>$GeneralWard\\\hspace*{0.6 in}$<$/User.locationAddress$>$\\
\hspace*{0.6 in}$<$User.requestTime rdf:datatype\\\hspace*{1 in}=``\&xsd;string"$>$DutyTime\\\hspace*{0.6 in}$<$/User.requestTime$>$\\
\hspace*{0.35 in}$<$/ComplexContext$>$}

\begin{figure}[h!]
\centering
\includegraphics[scale=0.70]{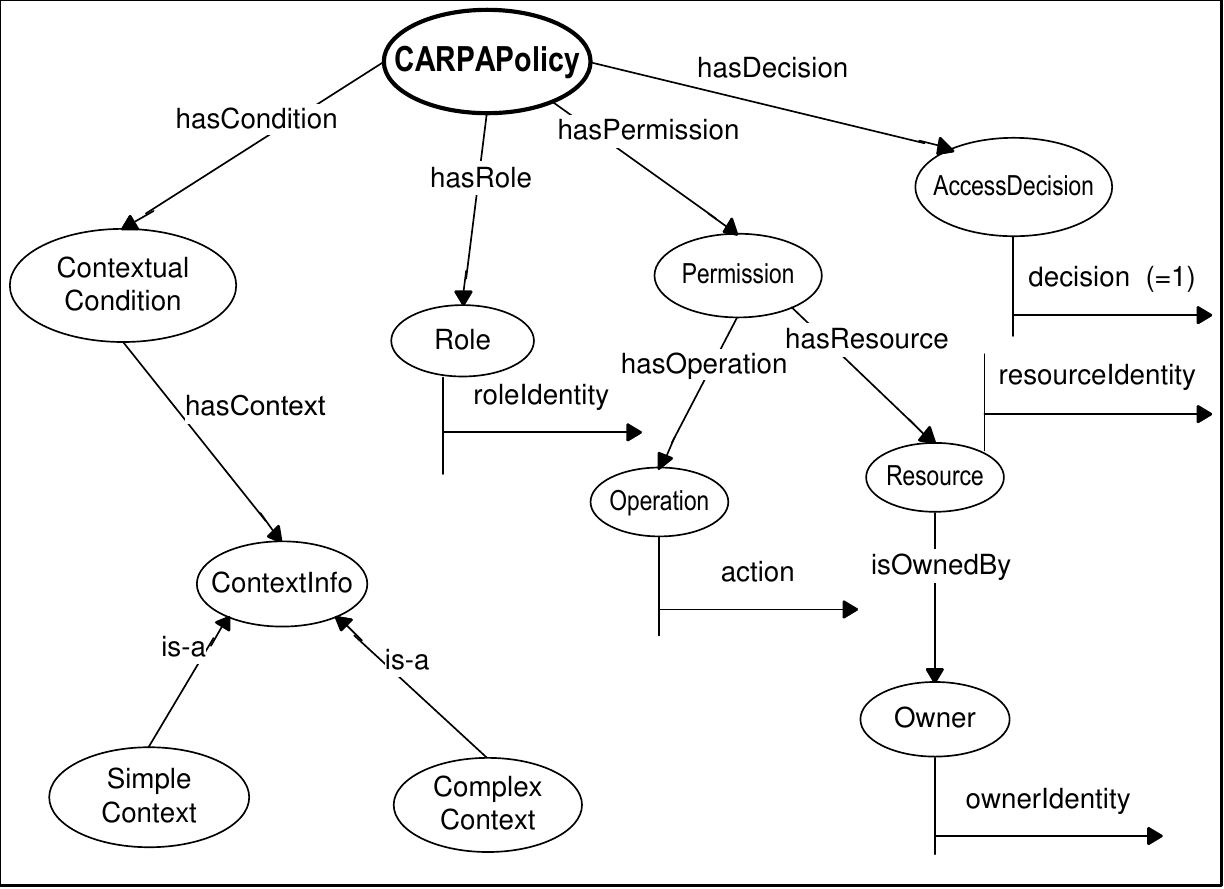}
\caption{The CARPA Policy Ontology}
\label{f:carpao}
\end{figure}

\begin{table*}
\begin{center}
\caption{Domain and Range Restrictions for CARPA Object Properties}
\vspace{-0.1in}
\label{drcarpa}
    \begin{tabular}{|p{3cm}|p{2.5cm}|p{2.5cm}|p{7.8cm}|}
    \hline
    \bf Object Property & \bf Domain & \bf Range & \bf Description \\ \hline \hline
hasDecision & CARPAPolicy & AccessDecision & A CARPA policy has the access decision \\ \hline
hasPermission & CARPAPolicy & Permission & A CARPA policy has a permission \\ \hline
hasOperation & Permission & Operation & A CARPA policy has a permission to access/perform different operations on resource \\ \hline
hasResource & Permission & Resource & A CARPA policy has a permission to access resource \\ \hline
isOwnedBy & Resource & Owner & A resource is owned by an owner \\ \hline
\end{tabular}
\end{center}
\end{table*}

\subsection{CARPA Policy Ontology}
\label{carpapo}

The CARPA policy ontology, representing {\it context-aware role-permission assignment policies}, has been designed by answering the following questions.
\begin{itemize}
\item Who is requesting access by playing what role (role)?
\item What type of object is being requested (resource or service)?
\item What is the dynamic context information that is relevant for this role-permission assignment (contextual condition)? 
\end{itemize}

\subsubsection{Core Concepts}
\vspace*{0.2 in}

A graphical representation of the ontology is shown in Figure \ref{f:carpao}. The ontology has the following core concepts, which are organized into a {\it CARPAPolicy} hierarchy, including such concepts as {\it Role}, {\it Permission}, {\it Resource}, {\it Operation}, {\it AccessDecision} and {\it ContextualCondition}. The {\it ContextualCondition} class has the concept {\it ContextInfo} which in turn has two subclasses {\it SimpleContext} and {\it ComplexContext}.

The {\it CARPAPolicy} ontology has {\it object} and {\it data type} properties. The domain and range of object properties are specified in Table \ref{drcarpa}.

The {\it data type} properties of the {\it CARPAPolicy} ontology are shown in Table \ref{dtpcarpa}.

A CARPA policy has exactly one access decision value (``Granted" or ``Denied"). To specify this cardinality constraint in our CARPA policy ontology (see Figure \ref{f:carpao}), we consider a class {\it AccessDecision}, and its data type property {\it decision}. The possible access decision values are summarized in Table \ref{carcon}.

\begin{table}
\begin{center}
\caption{CARPA Data Type Properties}
\vspace{-0.1in}
\label{dtpcarpa}
    \begin{tabular}{|p{3.8cm}|p{3.5cm}|}
    \hline
    \bf Data Type Property & \bf Domain \\ \hline \hline
decision & AccessDecision\\ \hline
resourceIdentity & Resource \\ \hline
ownerIdentity & Owner \\ \hline
action & Operation \\ \hline
\end{tabular}
\end{center}
\end{table}

\begin{table}
\begin{center}
\caption{Cardinality Constraint}
\label{carcon}
    \begin{tabular}{|p{1.3cm}|p{2.5cm}|p{3.4cm}|}
    \hline
    \bf Property & \bf Possible Values & \bf Description \\ \hline \hline
decision & Granted & Access request is granted \\ \hline
decision & Denied & Access request is denied \\ \hline

\end{tabular}
\end{center}
\end{table}

Overall, a CARPA policy captures the \textit{\textbf{3W}} dimensions which can be read as follows: 

\vspace*{0.1 in}
\begin{quote}
{\it a {\it CARPAPolicy} specifies that a user who is playing a {\it role} has {\it AccessDecision} (``Granted" or ``Denied") to which parts ({\it resourceIdentity}) of a {\it Resource} for a specific {\it action} (``Read" or ``Write" {\it Operation}) or a range of actions under what {\it contextual conditions}.} 
\end{quote}
\vspace*{0.1 in}

The details of OWL-based CARPA policy specifications can be found in Appendix B.

\begin{table*}
\begin{center}
\caption{An Example CARPA Policy for the Registered Nurse}
\vspace{-0.1in}
\label{table:rnrac}
    \begin{tabular}{|p{0.7cm}|p{12cm}|}
    \hline
1&\hspace*{0.1 in}$<${\bf CARPAPolicy} rdf:ID=``$carpa_{1}$"$>$\\
2&\hspace*{0.4 in}$<$hasRole rdf:resource=``\#Role\_RegisteredNurse"/$>$\\
3&\hspace*{0.4 in}$<$hasPermission rdf:resource=``\#Permission\_DMR\_Write"/$>$\\
4&\hspace*{0.4 in}$<$hasCondition rdf:resource=``\#ContextualCondition\_ContextInfo"/$>$\\
5&\hspace*{0.4 in}$<$hasDecision rdf:resource=``\#AccessDecision\_Granted"/$>$\\ 
6&\hspace*{0.1 in}$<${\bf /CARPAPolicy}$>$\\ \hline
7&\hspace*{0.1 in}$<${\bf Role} rdf:ID=``Role\_RegisteredNurse"$>$\\
8&\hspace*{0.4 in}$<$roleIdentity rdf:datatype=``\&xsd;string"$>$RN00X$<$/roleIdentity$>$\\
9&\hspace*{0.1 in}$<${\bf /Role}$>$\\ \hline
10&\hspace*{0.1 in}$<${\bf Permission} rdf:ID=``Permission\_DMR\_Write"$>$\\
11&\hspace*{0.4 in}$<$hasResource rdf:resource=``\#Resource\_DMR"/$>$\\
12&\hspace*{0.4 in}$<$hasOperation rdf:resource=``\#Operation\_Write"/$>$\\
13&\hspace*{0.1 in}$<${\bf /Permission}$>$\\ \hline
14& \hspace*{0.1 in}$<${\bf Resource} rdf:ID=``Resource\_DMR"$>$\\
15&\hspace*{0.4 in}$<$resourceIdentity rdf:datatype=``\&xsd;int"$>$2$<$/resourceIdentity$>$\\
16&\hspace*{0.1 in}$<${\bf /Resource}$>$\\ \hline
17&\hspace*{0.1 in}$<${\bf Operation} rdf:ID=``Operation\_Write"$>$\\
18&\hspace*{0.4 in}$<$action rdf:datatype=``\&xsd;string"$>$Write$<$/action$>$\\
19&\hspace*{0.1 in}$<${\bf /Operation}$>$ \\ \hline
20&\hspace*{0.1 in}$<${\bf AccessDecision} rdf:ID=``AccessDecision\_Granted"$>$\\
21&\hspace*{0.4 in}$<$decision rdf:datatype=``\&xsd;string"$>$Granted$<$/decision$>$\\
22&\hspace*{0.1 in}$<${\bf /AccessDecision}$>$\\ \hline
23&\hspace*{0.1 in}$<${\bf ContextualCondition} rdf:ID=``ContextualCondition\_ContextInfo"$>$\\
24&\hspace*{0.4 in}$<$hasContext rdf:resource=``\#ComplexContext\_$c_{c2}$"/$>$\\ 
25&\hspace*{0.1 in}$<${\bf /ContextualCondition}$>$
\\ \hline
\end{tabular}
\end{center}
\end{table*}

\subsubsection{An Example CARPA Policy}
\vspace*{0.2 in}
\begin{exmp}
\label{rnrpp}
Again consider the access control policy for the registered nurses (presented in Section \ref{rm}): a registered nurse, who is assigned for a regular follow-up visit to monitor a patient's health condition, can access the patient's daily medical records (DMR) when the patient's health status is normal. 
\end{exmp}

In this policy, the access decision is based on the following policy constraints: \textit{\textbf{who}} the user is (user's {\it role}), \textit{\textbf{what}} resource is being requested (resource's {\it identity}), and \textit{\textbf{under what contextual conditions}} the user sends the request (the {\it interpersonal relationship} between user and resource owner and the {\it health status} of the patient). The CARPA policy for the registered nurses in OWL is shown in Table \ref{table:rnrac}. The policy states that the registered nurse (the assigned nurse) can access the patient's daily medical records when the patient's health condition is normal. The core policy concepts are specified in {\it Line\# 1 to 6}, the role specification (registered nurse) is shown in {\it Line\# 7 to 9}, the permission specification (daily medical records on write operation) is shown in {\it Line\# 10 to 19}, and the access decision (granted decision) is specified in {\it Line\# 20 to 22}. The contextual condition (a complex context $c_{c2}$) is specified in {\it Line\# 23 to 25}.

The contextual condition $c_{c2}$ is already expressed in Formula (\ref{rncc2}), using logical composition (see Section \ref{csl}). Definition \ref{cc2owl} shows its ontological definition.

\begin{defn}
(`$c_{c2}$' Contextual Condition Definition).
\label{cc2owl}
\end{defn}
{\it \hspace*{0.2 in}$<$ComplexContext rdf:ID=``ComplexContext\_$c_{c2}$"$>$\\
\hspace*{0.5 in}$<$interRelationship(User, Owner) \\\hspace*{0.6 in}rdf:datatype=``\&xsd;string"$>$AssignedNurse\\\hspace*{0.5 in}$<$/interRelationship(User, Owner)$>$\\
\hspace*{0.5 in}$<$Owner.healthStatus \\ \hspace*{1 in}rdf:datatype=``\&xsd;string"$>$\\\hspace*{1 in}Normal$<$/Owner.healthStatus$>$\\
\hspace*{0.35 in}$<$/ComplexContext$>$}\\ 

The details of context information modelling can be found in the ontology-based context model, which has been presented in our earlier work \cite{KayesOnt13, KayesHC15CJ}.

\section{Policy Enforcement Architecture}
\label{pea}

This section introduces the policy enforcement architecture (PEA) of our framework and describes its components. PEA extends our previous implementation prototype, which is reported in \cite{KayesHC15CJ}.

\begin{figure}[t!]
\centering
\includegraphics[scale=0.75]{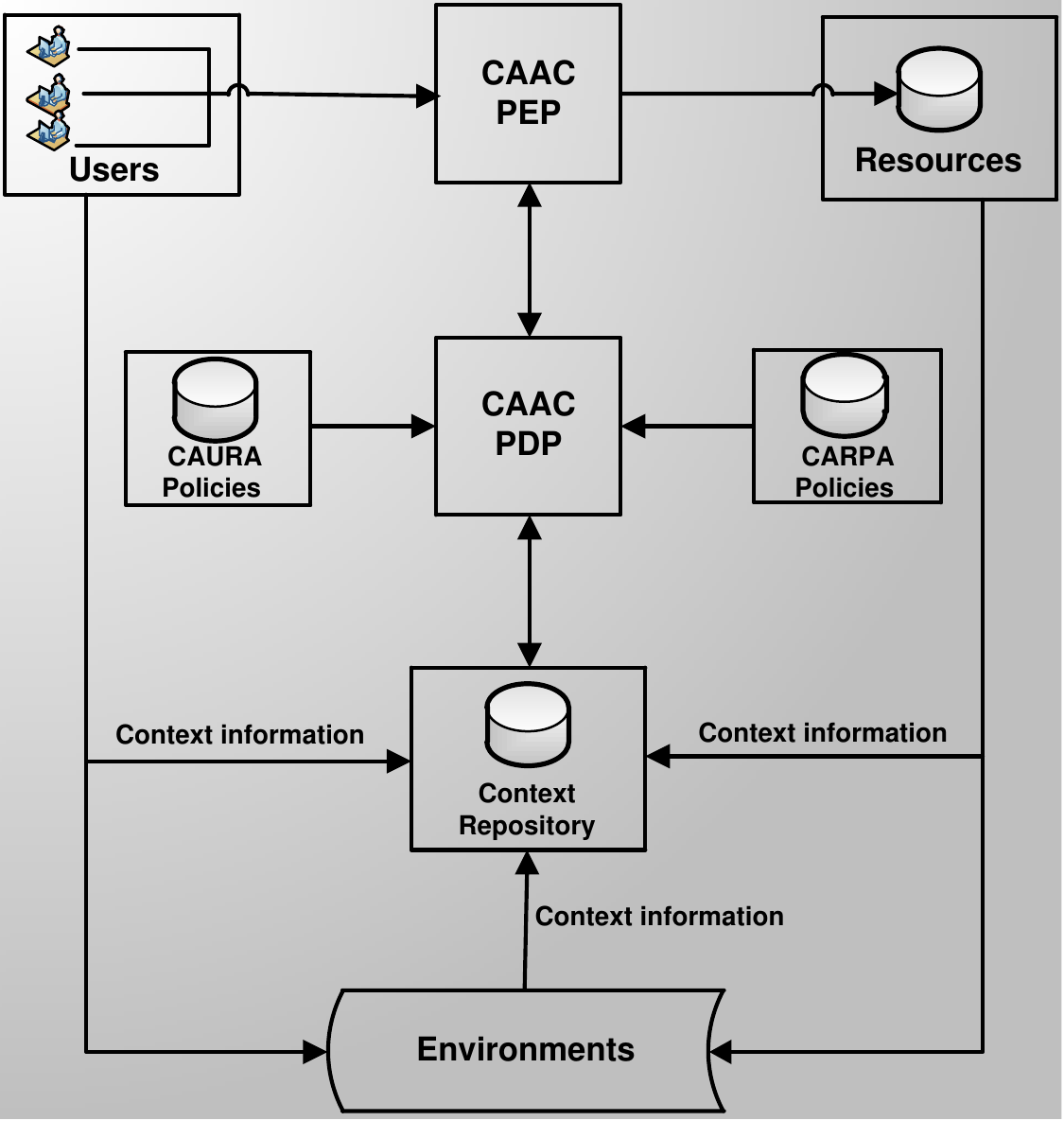}
\caption{The Policy Enforcement Architecture}
\label{figure:pea}
\end{figure}

Figure \ref{figure:pea} gives an overview of the PEA architecture. It includes four main components: Context Repository, CAAC Policies (CAURA and CARPA policies), CAAC PDP (policy decision point), and CAAC PEP (policy enforcement point).

The {\it Context Repository} stores the access control-specific context information in the form of a context ontology, including user-centric context information (e.g., requester profile), resource-centric context information (e.g., resource profile), and environment-centric context information (e.g., user's location, and interpersonal relationship between user and owner). Consequently, the contextual conditions for the user-role and role-permission assignments are specified in the {\it Context Repository} in terms of relevant context information. The detailed implementation of the context ontology for access control is the out of scope of this paper, which can be found in our earlier paper \cite{KayesOnt13, KayesHC15CJ}. 

The {\it CAURA Policy} and {\it CARPA Policy} ontologies store two sets of policies that show a mapping between users and roles, and roles and permissions respectively, according to the relevant contextual conditions that are in effect. We have used the Prot\'eg\'e-OWL API \cite{protege} to implement the context and policy ontologies. We have used the Jess Rule Engine \cite{jess} for executing the SWRL rules (user-defined reasoning rules) \cite{swrl}.

We have implemented the {\it CAAC PDP} and {\it CAAC PEP} in Java to evaluate the policies for making context-aware access control decisions. Once the {\it CAAC PEP} receives the user's request for resource access, it queries the {\it CAAC PDP} for the applicable policies and currently available context information in the ontologies. The {\it CAAC PDP} makes the decision according to the policies and context information. Finally, the {\it CAAC PDP} informs the {\it CAAC PEP} of the decision, and the {\it CAAC PEP} enforces the decision by granting or denying the user’s access request.

\section{The Evaluation of the Ontology-based Policy Model and Framework}
\label{acs}

In this section, we demonstrate the feasibility of our proposed ontology-based policy framework. We aim to show that the policy ontology model and framework offers complete and correct semantics and shows its efficiency in terms of response time. Our evaluation considers the following factors, namely, the {\it completeness} of the model components (ontology concepts), the {\it correctness} and {\it consistency} of the ontology semantics and the {\it performance} of the framework in terms of response time.

\subsection{Completeness}
\label{ecom}

First, we evaluate the completeness of our policy ontology model. As such, we in this section present the patient medical records management (PMRM) application, covering our framework features: contextual conditions, context-aware user role assignment (CAURA) policies and context-aware role-permission assignment (CARPA) policies. The main goal of this PMRM application is to control the users' access (read or write operation) to different medical records of patients based on the dynamic context information.

We present below two test cases, (i.e., the emergency and normal cases from our application scenario presented in Section \ref{rm}) that highlight specific policy requirements of the PMRM application, and demonstrate how the dynamic context information (contextual conditions) is incorporated into the CAURA and CARPA policies.

\subsubsection{Revisiting the Application Scenario - Emergency Case}

Consider the scenario when Jane (who is a general physician) wants to access the emergency medical records of patient Bob, an access request is submitted to the {\it CAAC PEP} (which is a part of the {\it Policy Enforcement Architecture (PEA)}) for evaluation. The {\it CAAC PEP} forwards the request to the {\it CAAC PDP}, which captures the applicable access control policies in the policy ontology. It also captures the relevant context information in the context repository. For Scene \#1, Jane's resource access request is shown as follows:\\

\hspace*{0.1 in}$<user = (Jane), 
\\\hspace*{0.6 in}permission = (EMR(Bob), write)>$\\

The core policy ontology captures the relevant policy constraints applicable to the user-role assignment: user-centric, resource-centric and environmental context information. The contextual condition $c_{s2}$ (shown in Table \ref{table:cc3}) is modeled/captured in our ontology, based on the available context information. It shows that the users can play the emergency doctors role from the emergency room of the hospital.

\begin{table}
\begin{center}
\caption{`$c_{s2}$' Contextual Condition Definition}
\vspace{-0.1in}
\label{table:cc3}
    \begin{tabular}{|p{7.8cm}|}
    \hline
$c_{s2} = (User.locationAddress = ``EmergencyRoom")$ \\
    \hline
\end{tabular}
\end{center}
\end{table}

\begin{table}
\begin{center}
\caption{An Example CAURA Policy for the Emergency Doctors}
\vspace{-0.1in}
\label{table:caura-ed}
    \begin{tabular}{|p{0.5cm}|p{7cm}|}
    \hline
1&\hspace*{0.05 in}$\textbf{\textit {If}}$\\
2&\hspace*{0.3 in}$CAURAPolicy(caura_{2})$\hspace*{0.05 in}$\wedge$\\
3&\hspace*{0.3 in}$User(u)$\hspace*{0.05 in}$\wedge$\hspace*{0.05 in}$hasUser(caura_{2}, u)$\hspace*{0.05 in}$\wedge$\\
4&\hspace*{0.3 in}$Role(r)$\hspace*{0.05 in}$\wedge$\hspace*{0.05 in}$hasRole(caura_{2}, r)$\hspace*{0.05 in}$\wedge$\\
5&\hspace*{0.3 in}$ContextualCondition(exp)$\hspace*{0.05 in}$\wedge$\\
6&\hspace*{0.3 in}$hasCondition(caura_{2}, exp)$\hspace*{0.05 in}$\wedge$\hspace*{0.05 in}\\
7&\hspace*{0.3 in}$has(u, userIdentity)$\hspace*{0.05 in}$\wedge$\\
8&\hspace*{0.3 in}$equal(userIdentity, ``doctorIdentity")$ \hspace*{0.05 in}$\wedge$\\
9&\hspace*{0.3 in}$has(r, roleIdentity)$\hspace*{0.05 in}$\wedge$\\
10&\hspace*{0.3 in}$equal(roleIdentity, ``ED00X")$\hspace*{0.05 in}$\wedge$\\
11&\hspace*{0.3 in}$equal(exp, ``c_{s2}")$ \\
12&\hspace*{0.05 in}$\textbf{\textit {Then}}$\\
13&\hspace*{0.3 in}$caura(u, r)$\\
    \hline
\end{tabular}
\end{center}
\vspace{-0.2in}
\end{table}

Table \ref{table:caura-ed} shows the context-aware user-role assignment (CAURA) policy for emergency doctors. The policy states that the users can play the emergency doctor role when they are present in the emergency room of the hospital. In this policy, the access decision is based on the following policy constraints: \textit{\textbf{who}} the user is (user's {\it identity}), \textit{\textbf{what}} role the user can play (role's {\it identity}) and \textit{\textbf{under what contextual condition}} (the {\it location} of the user). The CAURA policy for the emergency doctors is shown in Table \ref{table:caura-ed}. The core policy concepts are specified in {\it Line\# 1 to 6}, the user specification is shown in {\it Line\# 7 to 8}, the role specification (emergency doctor) is shown in {\it Line\# 9 to 10}, the contextual condition ($c_{s2}$) is specified in {\it Line\# 11} and the context-aware user-role assignment ($caura$) is specified in {\it Line\# 12 to 13}.

The policy ontology also captures the relevant policy constraints applicable to the role-permission assignment. Table \ref{table:cc4} expresses the contextual condition ($c_{s3}$) according to the relevant context information (the {\it health status} of the patient).

Table \ref{table:carpa-ed} presents the context-aware role-permission assignment (CARPA) policy for emergency doctors. It states that the emergency doctors can access the patient's emergency medical records when the patient's health condition is critical. In this policy, the access decision is based on the following policy constraints: \textit{\textbf{who}} the user is (user's {\it role}), \textit{\textbf{what}} resource is being requested (resource's {\it identity}) and \textit{\textbf{under what contextual condition}} the user sends the request (the {\it health status} of the patient). The CARPA policy for the emergency doctors is shown in Table \ref{table:carpa-ed}. The core policy concepts are specified in {\it Line\# 1 to 7}, the role specification (emergency doctor) is shown in {\it Line\# 8 to 9}, the permission specification (emergency medical records on write operation) is shown in {\it Line\# 10 to 14}, the contextual condition ($c_{s3}$) is specified in {\it Line\# 15} and the context-aware role-permission assignment ($carpa$) is specified in {\it Line\# 16 to 17}.

\begin{table}
\begin{center}
\caption{`$c_{s3}$' Contextual Condition Definition}
\vspace{-0.1in}
\label{table:cc4}
    \begin{tabular}{|p{7.8cm}|}
    \hline
$c_{s3} = (Owner.healthStatus = ``Critical")$ \\
    \hline
\end{tabular}
\end{center}
\end{table}

\begin{table}
\begin{center}
\caption{An Example CARPA Policy for the Emergency Doctors}
\vspace{-0.1in}
\label{table:carpa-ed}
    \begin{tabular}{|p{0.5cm}|p{7cm}|}
    \hline
1&\hspace*{0.05 in}$\textbf{\textit {If}}$\\
2&\hspace*{0.3 in}$CARPAPolicy(carpa_{2})$\hspace*{0.05 in}$\wedge$\\
3&\hspace*{0.3 in}$Role(r)$\hspace*{0.05 in}$\wedge$\hspace*{0.05 in}$hasRole(carpa_{2}, r)$\hspace*{0.05 in}$\wedge$\\
4&\hspace*{0.3 in}$Permission(p)$\hspace*{0.05 in}$\wedge$\\
5&\hspace*{0.3 in}$hasPermission(carpa_{2}, p)$\hspace*{0.05 in}$\wedge$\\
6&\hspace*{0.3 in}$ContextualCondition(exp)$\hspace*{0.05 in}$\wedge$\\
7&\hspace*{0.3 in}$hasCondition(carpa_{2}, exp)$\hspace*{0.05 in}$\wedge$\hspace*{0.05 in}\\
8&\hspace*{0.3 in}$has(r, roleIdentity)$\hspace*{0.05 in}$\wedge$\\
9&\hspace*{0.3 in}$equal(roleIdentity, ``ED00X")$\hspace*{0.05 in}$\wedge$\\
10&\hspace*{0.3 in}$Resource(res)$\hspace*{0.05 in}$\wedge$\hspace*{0.05 in}$hasResource(p, res)$\hspace*{0.05 in}$\wedge$\\
11&\hspace*{0.3 in}$Operation(op)$\hspace*{0.05 in}$\wedge$\hspace*{0.05 in}$hasOperation(p, op)$\hspace*{0.05 in}$\wedge$\hspace*{0.05 in}\\
12&\hspace*{0.3 in}$has(res, resourceIdentity)$\hspace*{0.05 in}$\wedge$\\
13&\hspace*{0.3 in}$equal(resourceIdentity, 1)$\hspace*{0.05 in}$\wedge$\\
14&\hspace*{0.3 in}$has(op, action)$\hspace*{0.05 in}$\wedge$\hspace*{0.05 in}$equal(action, ``write")$\hspace*{0.05 in}$\wedge$\\
15&\hspace*{0.3 in}$equal(exp, ``c_{s3}")$\\
16&\hspace*{0.05 in}$\textbf{\textit {Then}}$\\
17&\hspace*{0.3 in}$carpa(r, p)$\\
    \hline
\end{tabular}
\end{center}
\vspace{-0.2in}
\end{table}

Based on these CAURA and CARPA policies (Tables \ref{table:caura-ed} and \ref{table:carpa-ed}), the {\it CAAC PDP} (Figure \ref{figure:pea}) determines whether the request is ``granted" or ``denied" for the submitted access request, and returns the access decision to the {\it CAAC PEP}. Finally, the {\it CAAC PEP} enforces the context-aware access control decision, based on the applicable policies and the relevant contextual conditions. If the decision is ``granted", the requested resource is sent to the user; otherwise, a ``denied" response is sent to the user. For the application example ({\it Scene \#1}), Jane's resource access permission (``{\it EMR\_write}") is $granted$. 

This case study for the test scenario shows that our policy framework is able to successfully make access control decisions through context-aware user-role and role-permission assignments. In this scenario, Jane is not an emergency doctor, but he can play the emergency doctor role from the emergency room of the hospital and is allowed to access Bob's emergency medical records in such a critical situation. On the other hand, when the context changes (e.g., Jane leaves the emergency room, or Bob's health condition changes from critical to normal), the system will not grant Jane the access to the requested resource. In general, at each time of an access request or when context changes, the {\it CAAC PEP} sends automated request to the {\it CAAC PDP} for the applicable policies and the relevant context information.

\subsubsection{Revisiting the Application Scenario - Normal Case}

For Scene \#2 in our application scenario, where a registered nurse Mary wants to access Bob's daily medical records (DMR), an access request is submitted to the {\it CAAC PEP} for evaluation. Mary's resource access request is shown as follows:\\

\hspace*{0.1 in}$<user = (Mary), 
\\\hspace*{0.6 in}permission = (DMR(Bob), write)>$\\

Table \ref{table:caura-rn1} shows the CAURA policy to play the registered nurse role. The policy states that a user having a nurse identity can play the registered nurse role from the general ward of the hospital during her ward shift time. The specification of the contextual condition associated with this policy, named $c_{c1}$ (during ward duty time from the general ward of the hospital), is expressed in Formula (\ref{rncc1}), using logical composition (see Section \ref{csl}).

\begin{table}
\begin{center}
\caption{An Example CAURA Policy for the Registered Nurses}
\vspace{-0.1in}
\label{table:caura-rn1}
    \begin{tabular}{|p{0.5cm}|p{7cm}|}
    \hline
1&\hspace*{0.05 in}$\textbf{\textit {If}}$\\
2&\hspace*{0.3 in}$CAURAPolicy(caura_{1})$\hspace*{0.05 in}$\wedge$\\
3&\hspace*{0.3 in}$User(u)$\hspace*{0.05 in}$\wedge$\hspace*{0.05 in}$hasUser(caura_{1}, u)$\hspace*{0.05 in}$\wedge$\\
4&\hspace*{0.3 in}$Role(r)$\hspace*{0.05 in}$\wedge$\hspace*{0.05 in}$hasRole(caura_{1}, r)$\hspace*{0.05 in}$\wedge$\\
5&\hspace*{0.3 in}$ContextualCondition(exp)$\hspace*{0.05 in}$\wedge$\\
6&\hspace*{0.3 in}$hasCondition(caura_{1}, exp)$\hspace*{0.05 in}$\wedge$\hspace*{0.05 in}\\
7&\hspace*{0.3 in}$has(u, userIdentity)$\hspace*{0.05 in}$\wedge$\\
8&\hspace*{0.3 in}$equal(userIdentity, ``nurseIdentity")$ \hspace*{0.05 in}$\wedge$\\
9&\hspace*{0.3 in}$has(r, roleIdentity)$\hspace*{0.05 in}$\wedge$\\
10&\hspace*{0.3 in}$equal(roleIdentity, ``RN00X")$\hspace*{0.05 in}$\wedge$\\
11&\hspace*{0.3 in}$equal(exp, ``c_{c1}")$\\
12&\hspace*{0.05 in}$\textbf{\textit {Then}}$\\
13&\hspace*{0.3 in}$caura(u, r)$\\
    \hline
\end{tabular}
\end{center}
\end{table}

\begin{table}

\begin{center}
\caption{An Example CARPA Policy for the Registered Nurses}
\vspace{-0.1in}
\label{table:carpa-rn1}
    \begin{tabular}{|p{0.5cm}|p{7cm}|}
    \hline
1&\hspace*{0.05 in}$\textbf{\textit {If}}$\\
2&\hspace*{0.3 in}$CARPAPolicy(carpa_{1})$\hspace*{0.05 in}$\wedge$\\
3&\hspace*{0.3 in}$Role(r)$\hspace*{0.05 in}$\wedge$\hspace*{0.05 in}$hasRole(carpa_{1}, r)$\hspace*{0.05 in}$\wedge$\\
4&\hspace*{0.3 in}$Permission(p)$\hspace*{0.05 in}$\wedge$\\
5&\hspace*{0.3 in}$hasPermission(carpa_{1}, p)$\hspace*{0.05 in}$\wedge$\\
6&\hspace*{0.3 in}$ContextualCondition(exp)$\hspace*{0.05 in}$\wedge$\\
7&\hspace*{0.3 in}$hasCondition(carpa_{1}, exp)$\hspace*{0.05 in}$\wedge$\hspace*{0.05 in}\\
8&\hspace*{0.3 in}$has(r, roleIdentity)$\hspace*{0.05 in}$\wedge$\\
9&\hspace*{0.3 in}$equal(roleIdentity, ``RN00X")$\hspace*{0.05 in}$\wedge$\\
10&\hspace*{0.3 in}$Resource(res)$\hspace*{0.05 in}$\wedge$\hspace*{0.05 in}$hasResource(p, res)$\hspace*{0.05 in}$\wedge$\\
11&\hspace*{0.3 in}$Operation(op)$\hspace*{0.05 in}$\wedge$\hspace*{0.05 in}$hasOperation(p, op)$\hspace*{0.05 in}$\wedge$\hspace*{0.05 in}\\
12&\hspace*{0.3 in}$has(res, resourceIdentity)$\hspace*{0.05 in}$\wedge$\\
13&\hspace*{0.3 in}$equal(resourceIdentity, 2)$\hspace*{0.05 in}$\wedge$\\
14&\hspace*{0.3 in}$has(op, action)$\hspace*{0.05 in}$\wedge$\hspace*{0.05 in}$equal(action, ``write")$\hspace*{0.05 in}$\wedge$\\
15&\hspace*{0.3 in}$equal(exp, ``c_{c2}")$\\
16&\hspace*{0.05 in}$\textbf{\textit {Then}}$\\
17&\hspace*{0.3 in}$carpa(r, p)$\\
    \hline
\end{tabular}
\end{center}
\end{table}

The CARPA policy for the registered nurses is shown in Table \ref{table:carpa-rn1}. The policy states that a registered nurse, who is assigned for a regular follow-up visit to monitor a patient's health condition, can access the patient's daily medical records (resource identity is 2) when the patient's health condition is normal. The specification of the contextual condition $c_{c2}$ (the nurse is assigned for the patient and the patient's health status is normal) is expressed in Formula (\ref{rncc2}) in Section \ref{csl}.

Based on the above CAURA and CARPA policies (Tables \ref{table:caura-rn1} and \ref{table:carpa-rn1}), we can observe that Mary can play the registered nurse role if she is located in the general ward during her ward shift time and consequently, she is authorized to access the daily medical records of patient Bob, who is hosted in that ward in his normal health condition.

For the same scenario (Scene \#2), let us consider another access control policy: a registered nurse, who is assigned to monitor the patient's health condition, can access the patient's private medical records (PMR) if she is present with the patient. Mary's resource access request is shown as follows:\\

\hspace*{0.1 in}$<user = (Mary), 
\\\hspace*{0.6 in}permission = (PMR(Bob), read)>$\\

The same CAURA policy specified in Table \ref{table:caura-rn1} can be used for the user-role assignment. It states that Mary can play the registered nurse role from the general ward of the hospital during her ward duty time.

\begin{table}
\begin{center}
\caption{An Example CARPA Policy for the Registered Nurses}
\vspace{-0.1in}
\label{table:carpa-rn2}
    \begin{tabular}{|p{0.5cm}|p{7cm}|}
    \hline
1&\hspace*{0.05 in}$\textbf{\textit {If}}$\\
2&\hspace*{0.3 in}$CARPAPolicy(carpa_{3})$\hspace*{0.05 in}$\wedge$\\
3&\hspace*{0.3 in}$Role(r)$\hspace*{0.05 in}$\wedge$\hspace*{0.05 in}$hasRole(carpa_{1}, r)$\hspace*{0.05 in}$\wedge$\\
4&\hspace*{0.3 in}$Permission(p)$\hspace*{0.05 in}$\wedge$\\
5&\hspace*{0.3 in}$hasPermission(carpa_{1}, p)$\hspace*{0.05 in}$\wedge$\\
6&\hspace*{0.3 in}$ContextualCondition(exp)$\hspace*{0.05 in}$\wedge$\\
7&\hspace*{0.3 in}$hasCondition(carpa_{1}, exp)$\hspace*{0.05 in}$\wedge$\hspace*{0.05 in}\\
8&\hspace*{0.3 in}$has(r, roleIdentity)$\hspace*{0.05 in}$\wedge$\\
9&\hspace*{0.3 in}$equal(roleIdentity, ``RN00X")$\hspace*{0.05 in}$\wedge$\\
10&\hspace*{0.3 in}$Resource(res)$\hspace*{0.05 in}$\wedge$\hspace*{0.05 in}$hasResource(p, res)$\hspace*{0.05 in}$\wedge$\\
11&\hspace*{0.3 in}$Operation(op)$\hspace*{0.05 in}$\wedge$\hspace*{0.05 in}$hasOperation(p, op)$\hspace*{0.05 in}$\wedge$\hspace*{0.05 in}\\
12&\hspace*{0.3 in}$has(res, resourceIdentity)$\hspace*{0.05 in}$\wedge$\\
13&\hspace*{0.3 in}$equal(resourceIdentity, 3)$\hspace*{0.05 in}$\wedge$\\
14&\hspace*{0.3 in}$has(op, action)$\hspace*{0.05 in}$\wedge$\hspace*{0.05 in}$equal(action, ``read")$\hspace*{0.05 in}$\wedge$\\
15&\hspace*{0.3 in}$equal(exp, ``c_{c3}")$\\
16&\hspace*{0.05 in}$\textbf{\textit {Then}}$\\
17&\hspace*{0.3 in}$carpa(r, p)$\\
    \hline
\end{tabular}
\end{center}
\end{table}

\begin{table}
\begin{center}
\caption{`$c_{c3}$' Contextual Condition Definition}
\vspace{-0.1in}
\label{table:cc5}
    \begin{tabular}{|p{7.86cm}|}
    \hline
$c_{c3} = ((Owner.healthStatus = ``Normal")  \hspace*{0.05 in}\wedge$\\
$\hspace*{0.6 in}(interRelationship(User, Owner)$\\
$ \hspace*{1.7 in}= ``AssignedNurse") \hspace*{0.05 in}\wedge$\\
$\hspace*{0.6 in}(locationCentricRelationship(User, Owner)$\\
\hspace*{1.7 in}$ = ``Colocated"))$ \\
    \hline
\end{tabular}
\end{center}
\vspace{-0.1in}
\end{table}

The CARPA policy for the registered nurse is shown in Table \ref{table:carpa-rn2}. The policy states that a registered nurse can access the patient's private medical records (resource identity is 3), satisfying the contextual condition $c_{c3}$. The specification of $c_{c3}$ (the nurse is assigned to monitor the patient's health condition, and they both are co-located and the patient's health status is normal) is expressed  in Table \ref{table:cc5}.

Based on the CAURA and CARPA policies (Tables \ref{table:caura-rn1} and \ref{table:carpa-rn2}), we can observe that Mary can play the registered nurse role if she is located in the general ward during her ward shift time and consequently, she is authorized to access the private medical records of patient Bob, who is hosted in that ward in his normal health condition (as she is assigned to regularly monitor the health condition of the patient Bob). 

The above-mentioned healthcare case study for the two test scenarios guarantees the {\it completeness} of the policy ontology model. We observe that the CAAC policies for the two application scenarios are successfully specified by instantiating the domain-specific ontologies with the core policy ontology. Thus, the completeness of the policy ontology through presenting a case study from the healthcare domain shows the applicability of our policy framework.

\begin{table}
\begin{center}
\caption{Query 1}
\vspace{-0.1in}
\label{table:addpol}
 \begin{tabular}{|p{8cm}|}
    \hline
User({\bf ?user}) $\wedge$ Role({\bf ?role}) $\wedge$ Operation({\bf ?action}) $\wedge$ AccessDecision({\bf ?decision}) $\rightarrow$ {\bf sqwrl:select}(?user, ?role, ?action, ?decision) \\ \hline
\end{tabular}
\end{center}
\vspace{-0.1in}
\end{table}

\begin{table}
\begin{center}
\caption{Query 1 result}
\vspace{-0.1in}
\label{table:addpolr}
   \begin{tabular}{|p{1.2cm}|p{1.2 cm}|p{2.15cm}| p{2.15cm}|}
    \hline
  \bf ?user & \bf ?role & \bf ?action & \bf ?decision \\ \hline
 Tom & GR & Read & Granted \\
\hline
\end{tabular}
\end{center}
\vspace{-0.1in}
\end{table}

\subsection{Correctness and Consistency}

Second, we assess the correctness of the policy ontology \cite{FudholiRP15}. As such, we add some new domain-specific concepts into the ontology and specify the relevant context-aware access control policies. Also, we delete some ontology concepts. To identify the relevant changes, we execute some SQWRL queries \cite{sqwrl}. Finally, we verify these query results to evaluate the semantic correctness of the policy ontology. To this extent, we also assess the possible consistency of the ontology.

We have added some domain-specific concepts (individuals and attribute values) into the ontology and specified relevant CAAC policy rules (i.e., CAURA and CARPA policies). Then, we have executed some SQWRL queries to retrieve the query results starting with the empty query. For example, we have added a new {\it Role} named {\it GuestResearcher} (simply, {\it GR}) and an individual named Tom (who is an instance of {\it GuestResearcher}) to the role ontology. The researchers are not directly healthcare members but they may need to access some of the patient information. The details of role ontology can be found in our earlier work \cite{KayesHC15CJ}. Also, we have specified the relevant CAAC policies for guest researchers. The user-role and role-permission policy specifications are already discussed in Section \ref{ecom}.

\begin{table}
\begin{center}
\caption{Query 2}
\vspace{-0.1in}
\label{table:delpol}
 \begin{tabular}{|p{8cm}|}
    \hline
User({\bf ?user}) $\wedge$ Role({\bf ?role}) $\wedge$ Operation({\bf ?action}) $\wedge$ AccessDecision({\bf ?decision}) $\rightarrow$ {\bf sqwrl:select}(?user, ?role, ?action, ?decision) \\ \hline
\end{tabular}
\end{center}
\vspace{-0.1in}
\end{table}

\begin{table}
\begin{center}
\caption{Query 2 result}
\vspace{-0.1in}
\label{table:delpolr}
   \begin{tabular}{|p{1.2cm}|p{1.2 cm}|p{2.15cm}| p{2.15cm}|}
    \hline
  \bf ?user & \bf ?role & \bf ?action & \bf ?decision \\ \hline
 ? & ? & ? & ? \\
\hline
\end{tabular}
\end{center}
\vspace{-0.1in}
\end{table}

Table \ref{table:addpol} shows a SQWRL query to retrieve knowledge from policy ontology and Table \ref{table:addpolr} shows the query result. The result shows that Tom is authorized to access ({\it read} permission) the patient's medical records. For simplicity, the relevant contextual conditions and other policy constraints are not shown in Tables \ref{table:addpol} and \ref{table:addpolr}.

We have also deleted some domain-specific concepts from the ontology and executed some SQWRL queries to verify the changes. For example, all the {\it RegisteredNurse} instances are deleted from the ontology. Tables \ref{table:delpol} and \ref{table:delpolr} show one of the relevant SQWRL queries and the query result respectively. The result shows no output, that is there are no applicable CAAC policies for {\it RegisteredNurse}. When any concept has been deleted from the ontology, the implemented ontology-based policy framework automatically removes the redundant access control policies accordingly. This ensures the possible consistency for change requests that is required for a rule-based framework.

The query results in the above two tables demonstrate that our policy ontology contains correct semantic knowledge to instantiate the core ontology into its domain-specific ontologies. In other words, the policy ontology does not contain any conflicting and inconsistent information.

\subsection{Performance}
\label{exp}

In addition to the completeness, correctness and consistency, we assess the performance of our policy framework, where we measure the query response time to provide resource access permissions to users. We have conducted two sets of experiments with our framework as applied to the PMRM application on a Windows 8.1 operating system running on Intel(R) Core(TM) i7 @ 3.0 GHz processor with 8GB of memory.

\textbf{\textit {Test 1.}} The first test focuses on measuring the response time of context-aware user-role assignments (CAURA) in the light of increasing the number of policy rules. We first codify 50 policy rules that are attached to 20 different health professionals roles ({\it EmergencyDoctor}, {\it RegisteredNurse}) according to the contextual conditions. Then, the number of policies is varied from 50 to 500 in increment of 50 with 138 different health professional roles \cite{ASCO}. In the case study section (see Section \ref{acs}), we have already codified several CAURA policy rules. To measure the response time of CAURA assignments, the average value of the 10 execution runs is used for the analysis.  

\begin{figure}[t]
\begin{center}
\includegraphics[scale=0.45]{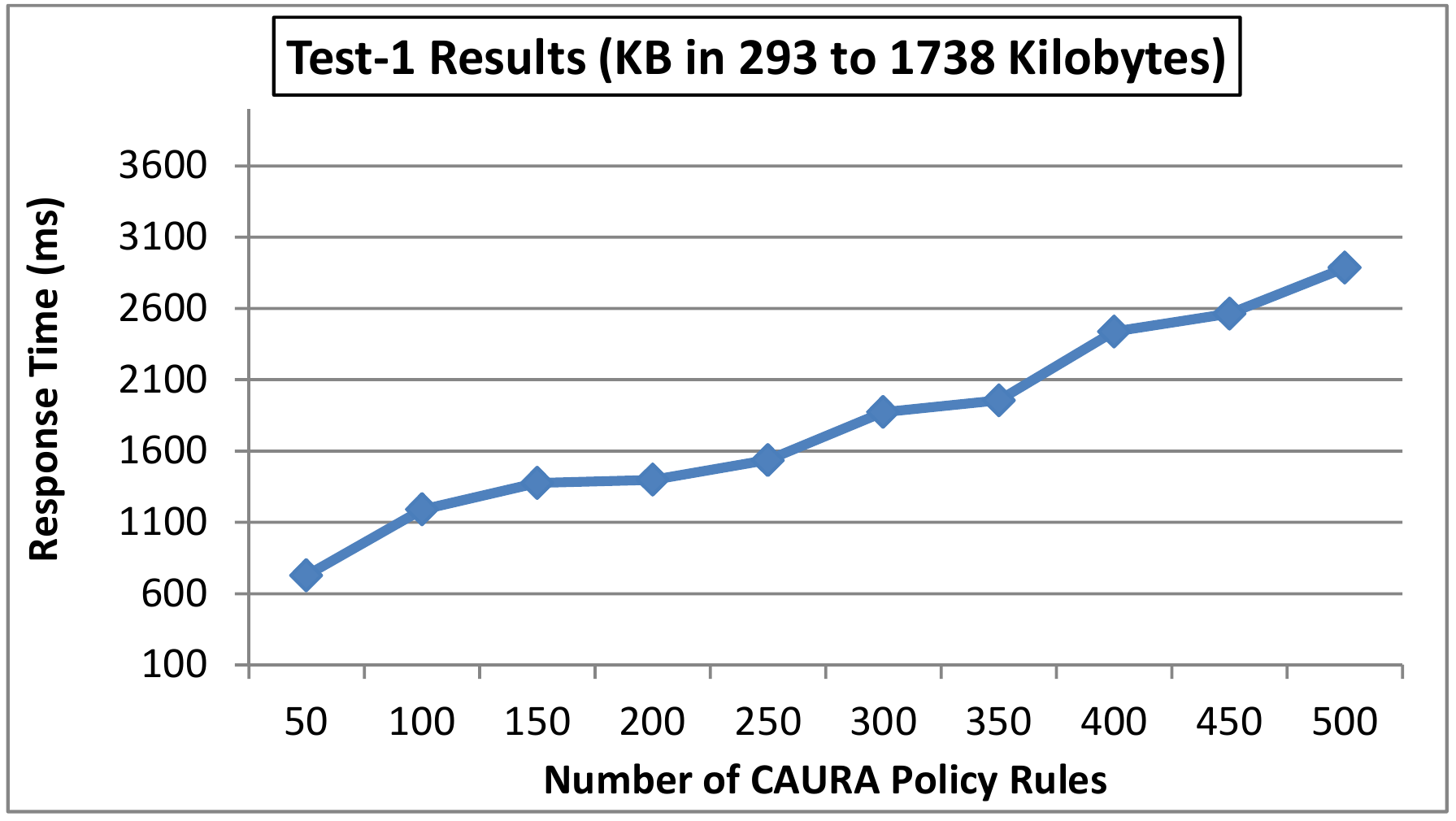}
 \end{center}
\vspace{-0.2in}
\caption{Response Time vs Number of Policies}
\label{fig:t1}
\end{figure}

The test results in Figure \ref{fig:t1} show that the average response time varies from 0.7 to 2.9 seconds approximately, as the number of CAURA policy rules changes and the size of the ontology KB (the context and policy ontologies) varies from 293 to 1738 Kilobytes respectively. We can see that the response time seems to be linear, relative to the number of policies.

\textbf{\textit {Test 2.}} In this test, we have evaluated the response time of context-aware role-permission assignments (CARPA) when the number of policy rules increased. Similar to Test 1, first, we have selected 50 CARPA policy rules with respect to 20 different health professional roles, then, we have varied the number of policies up to 500 in increment of 50 with the same 138 roles \cite{ASCO}. Each of these variations is executed 10 times for the analysis. In the case study section, we have already codified several CARPA policy rules.  

\begin{figure}[t]
\begin{center}
\includegraphics[scale=0.46]{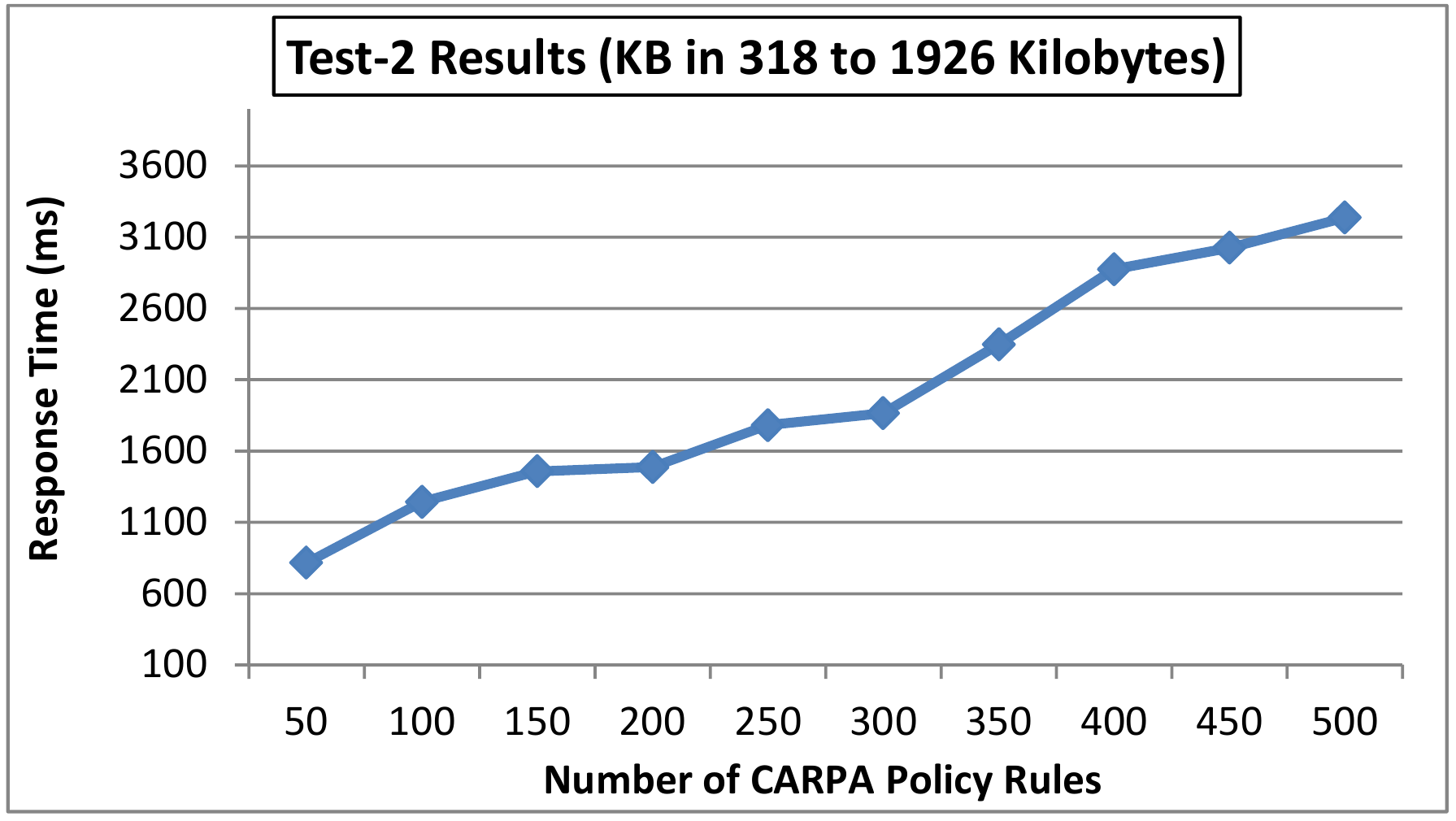}
 \end{center}
\vspace{-0.2in}
\caption{Response Time vs Number of Policies}
\label{fig:t2}
\end{figure}

The test results in Figure \ref{fig:t2} show that the response time increases when the number of policy rules increases. It varies between 0.8 and 3.2 seconds approximately, where the ontology KB size varies from 318 to 1926 Kilobytes respectively. Similar to test 1, in Figure \ref{fig:t2} we can see that the response time seems to be linear.  

In the above two tests, the computational overhead increases at a linear rate due to increasing the ontology KB sizes. At the point where we have specified 1000 policy rules (including both CAURA and CARPA policies), it takes approximately 6 seconds to process a user's request to access the resources. Overall, we consider that the {\it performance is acceptable} in such a setup with limited computing resources. That is, we can say that our framework has acceptable response time in supporting users' access to resources in a context-aware manner.

\section{Related Work and Comparative Analysis}
\label{rw}

In this section, we review the existing literature on role-based access control approaches in the context of the dynamically changing information (e.g., the location of the requesters) \cite{Dey01,DeyAS01}. Traditional role-based access control approaches \cite{Ferraiolo92,SandhuCFY96} exploit user identity/role information to determine the set of access permissions, whereas the dynamic context information can further limit the applicability of the available permissions. Our review includes the context-dependent role-based access control approaches that incorporate different types of context information into the traditional role-based access control (RBAC) process. We distinguish four different categories of access control approaches.

\begin{enumerate}[(i)]
\item Temporal role-based access control
\item Spatial role-based access control
\item Spatio-temporal role-based access control
\item Other context-dependent role-based access control 
\end{enumerate}

\subsection{Temporal Role-Based Access Control Approaches} 

The temporal role-based access control approaches \cite{BertinoBF01, JoshiBLG05} extend the basic role-based access control (RBAC) approaches \cite{Ferraiolo92,SandhuCFY96} by taking into account the temporal information (e.g., request times of users).

Bertino et al. \cite{BertinoBF01} have proposed the temporal RBAC (TRBAC) approach, which extends the traditional RBAC approach in order to support temporal constraints on enabling roles. To describe temporal constraints, TRBAC introduces a concept named role enabling base (REB), which is composed of periodic events and role triggers. For example, the periodic events and role triggers in the REB state that the {\it doctor-on-night-duty} role should be enabled during the night, whereas the role {\it doctor-on-day-duty} should be enabled during the day. TRBAC considers the temporal-aware user-role assignments, however, this approach does not provide adequate functionalities to integrate context information for role-permission assignments.

On the other hand, Joshi et al. \cite{JoshiBLG05} have extended the TRBAC approach proposed in \cite{BertinoBF01}. They proposed a generalized temporal role based access control (GTRBAC) model that allows specification of a comprehensive set of time-based access control policies, incorporating the temporal constraints in both user-role and role-permission assignments. 

These approaches take into account the temporal information when enforcing access control policies. However, they do not provide ontology-based implementation to realize the formal approaches, what we have in our policy framework. In our application scenario (presented in Section \ref{rm}), Mary should not assign to nurse role and consequently a nurse should not have access to medical records of the patients other than satisfying some contextual conditions (during ward duty time, from the hospital location, relationship between them are assigned nurse, etc.). Other than the temporal information, we also consider other relevant context information for user-role and role-permission assignments.

\subsection{Spatial Role-Based Access Control Approaches} 

The spatial role-based access control approaches \cite{BertinoCDP05, ZhangHS06} extend the basic role-based access control (RBAC) approaches \cite{Ferraiolo92,SandhuCFY96} by taking into account the spatial information (e.g., locations of users).

The GEO-RBAC \cite{BertinoCDP05} approach proposes the spatial extent (i.e., geographical location) of role, extending the traditional RBAC approach with the concept of spatial role. A spatial role represents a geographically bounded organizational function. The boundary specifies the spatial extent in which the user is located and enabled to play such a role. The GEO-RBAC approach provides spatial-aware user-role assignment and allows access to resources based on the spatial role the user holds within the spatial boundary. 

Zhang et al. \cite{ZhangHS06} have proposed a location-aware RBAC approach, named LRBAC. Like GEO-RBAC, LRBAC introduces the concept of spatial role. The roles are automatically activated/deactivated by the locations of the users. In LRBAC, both the activated roles of the users and their locations are taken into account for role-permission assignments, in order to evaluate the access control policies. 

These approaches take into account the spatial information when enforcing RBAC policies. Similar to above mentioned temporal RBAC approaches, however, they do not consider other relevant context information for user-role and role-permission assignments. Compared with these location-aware approaches, we have presented an ontology-based policy model in order to provide the practical basis for realizing the formal model.

\subsection{Spatio-Temporal Role-Based Access Control Approaches} 

The spatio-temporal role-based access control approaches \cite{ChandranJ05, BhattiGBJ05} extend the basic role-based access control (RBAC) approaches \cite{Ferraiolo92,SandhuCFY96} by taking into account both the spatial and temporal information.

Chandran and Joshi \cite{ChandranJ05} have extended the GTRBAC approach proposed in \cite{JoshiBLG05}. They have proposed a location and time-based RBAC approach, named LoT-RBAC \cite{ChandranJ05}. It considers temporal and spatial context information as contextual conditions. It adopts and extends the concept of role activation from the GTRBAC approach, by incorporating the temporal and spatial context information. In particular, a role is activated by a user from the location {\it l} at time {\it t}. LoT-RBAC allows access to resources if the location and temporal information of a user associated with the role activation is satisfied.

Bhatti et al. \cite{BhattiGBJ05} have proposed a spatio-temporal access control approach, named X-GTRBAC. It adopts the temporal-aware user-role and role-permission assignment policies from the GTRBAC approach, and incorporates spatial context information in these assignments.

These access control approaches consider the spatial and temporal information when enforcing RBAC policies. They have similar drawbacks in considering only the specific types (temporal and spatial) of context information for user-role and role-permission assignments. Also, they lack in providing a practical approach to incorporate the relevant context information into RBAC process.

\subsection{Other Context-Dependent Role-Based Access Control Approaches} 

Over the last few years, there are several research works extend the basic role-based access control (RBAC) approaches \cite{Ferraiolo92,SandhuCFY96}, where authorizations to access resources are based on the user assigned role and the relevant context information.

Kulkarni and Tripathi \cite{KulkarniT08} have proposed a context-aware role-based access control (CA-RBAC) model for pervasive computing applications. Using this model, they also present a programming framework for building context-aware access control applications. They consider user and resource-centric attributes as the context conditions in role-permission assignments. A user having a role and by fulfilling those conditions can access the resources. In contrast to the CA-RBAC model, the mapping of users to roles in our model is dynamically performed in accordance with the relevant context information. Different from this model, we also present a formal model to specify two sets of context-aware access control policies. In addition, our model includes a language to express contextual conditions based on the simple and complex context information.

Wang et al. \cite{WangLF08} have proposed a context-aware environment-role-based access control (CERBAC) approach for web services. They consider subject roles and environment roles as the access conditions. The policy rules are executed at runtime to grant or deny access based on both the subject roles and the environment roles. As such, access control decisions in CERBAC not only depend on the subject roles but also on environment roles. In CERBAC, the environment conditions are specified and modeled by the context information and they are used to define the environment roles. The unification of all relevant states (subject and environment states) into a single concept (roles) makes access control policies significantly easier to define and implement. However, the approach is not suitable for context-aware environments, because of the many roles (especially, contextual or environment roles) making the system very hard to maintain. In addition, it does not consider the context-aware user-role assignments.

A dynamic role-based access control (DRBAC) approach that incorporates the required credentials of users as context information when making user-role and role-permission assignments \cite{ZhengZZT11}. DRBAC only presents the concepts and requirements of the dynamic access control, without providing context and policy modelling supports. Compared with this work, we have presented both the formal and ontology-based policy models for our context-aware access control framework with an implementation (software) architecture and prototype. 

Kr{\"o}tzsch et al. \cite{H11} have considered access control for web service based on the user role and presented a policy model. This model is limited to considering specific types of contexts as policy constraints. Similar to the above approaches, this model has limitation in considering dynamic user-role and role-permission assignments in accordance with basic RBAC process. 

Schefer-Wenzl and Strembeck \cite{ScheferWenzlS13} have proposed an approach to context-aware role-based access control in ubiquitous environments. They propose a formal meta model that extends the UML, to integrate context attributes into the basic RBAC process. In this approach, the context attributes represent a certain properties of the environment such as time and location. It allows access to resources if this environment context information of a user associated with the role-permission assignments are satisfied. However, this context-aware role-based access control approach does not provide adequate functionalities to specify context-aware user-role assignments. In addition, the approach has limitation in providing a formal policy model to specify the two sets of context-aware (user-role and role-permission) access control policies. 

Kayes et al. \cite{KayesOnt13, KayesHC15CJ} have proposed an ontology-based approach to context-aware access control for information resources. They propose a context model for capturing and reasoning about access control-specific dynamic contexts, and a policy model for specifying and enforcing context-aware access control (CAAC) policies. This policy model extends the basic RBAC model, including the dynamic assignments of permissions to users based on the relevant contexts. However, this approach does not consider the context-aware user-role assignments. Also, it lacks in providing a formal policy model to specify and incorporate the relevant context information into both the user-role and role-permission assignments. Different from this approach, our policy framework also includes a simple language named context specification language for expressing contextual conditions based on the simple and complex context information.

Another recent ontological framework for situation-aware access control of software services has been proposed by Kayes et al. \cite{KayesHC14CAISE, KayesHC15IS}. The framework includes a situation model for identifying the relevant purpose-oriented situations and a policy model for specifying and enforcing situation-aware access control policies. The access control policies are specified in accordance with the possible situations. This policy model extends the concept of common role-permission assignment in RBAC, incorporating the concept of purpose-oriented situation. However, the framework does not provide adequate concepts to model both the context-aware user-role and role-permission assignment policies. Also, it lacks in providing a language to specify the different types of context expressions based on the simple and complex contexts.

Recently, Hosseinzadeh et al. \cite{HosseinzadehVRL16} and Trnka and Cern{\'{y}} \cite{TrnkaC16} have proposed the context-aware role-based access control models. In \cite{HosseinzadehVRL16}, roles are assigned to the users by the administrators and at runtime, users can access the resources based on the roles and the context information. The context of location and time are taken into account in deciding the restrictions. For example, in the healthcare domain a doctor is restricted to only read the medical history of the patients after office time or outside the hospital. In \cite{TrnkaC16}, the access control decisions are based on the context information (e.g., the range of IP addresses and times of the day) in addition to traditional roles in RBAC. In these two models, instead of considering two sets of context-aware user-role and role-permission assignment policies, they consider the static user-role assignment policies created by the administrator and the role-permission assignment policies in accordance with the context information. They also lack in providing a formal context-aware access control policy model. 

\vspace{0.1in}
\subsection{Comparative Analysis and Discussion}

This section presents the comparative analysis of the existing context-dependent RBAC approaches. The comparison is carried out in three groups: (i) the spatial and temporal-aware RBAC approaches, (ii) other context-specific RBAC approaches, and (iii) our earlier context/situation-aware RBAC approaches, with respect to the policy framework presented in this paper. 


\afterpage{
   \clearpage
   \thispagestyle{empty}
    \begin{landscape}
        \centering 

\begin{table*}
\begin{center}
\caption{Comparative analysis of the spatial and temporal RBAC approaches}
\vspace{-0.1in}
\label{table:ca-rbac}
\hspace*{-2.5 in}
\begin{tabular}{|c|c|c|c|c|c|c|c|c|c|c|}
\hline
	\multicolumn{1}{|p{5 cm}||}{\multirow{2}{*} {\bf RBAC Frameworks}} &
		\multicolumn{3}{c||}{\bf Formal Model} &
		\multicolumn{3}{c||}{\bf Implementation Model} &
		\multicolumn{4}{c|}{\bf Evaluation}\\
	\cline{2-11}
	\multicolumn{1}{|p{5cm}||}{\multirow{2}{*} {}} &
		\multicolumn{1}{p{1.3cm}|}{\bf CAURA Policy} & \multicolumn{1}{p{1.3cm}|}{\bf CARPA Policy} & 
		\multicolumn{1}{p{0.7cm}||}{\bf CSL} &
		\multicolumn{1}{p{1.5cm}|}{\bf CAURA Ontology} & \multicolumn{1}{p{1.5cm}|}{\bf CARPA Ontology} & 
		\multicolumn{1}{p{1.8cm}||}{\bf Software Prototype} &
		\multicolumn{1}{p{2.2cm}|}{\bf Completeness} & \multicolumn{1}{p{2cm}|}{\bf Correctness} & 
		\multicolumn{1}{p{2cm}|}{\bf Consistency} & \multicolumn{1}{p{2cm}|}{\bf Performance} \\
\hline
		\multicolumn{1}{|p{5cm}||}{TRBAC \cite{BertinoBF01}} &
		\multicolumn{1}{p{1.3cm}|}{$P/A$} & \multicolumn{1}{p{1.3cm}|}{$N/A$} & 
		\multicolumn{1}{p{0.7cm}||}{$N/A$} &
		\multicolumn{1}{p{1.5cm}|}{$N/A$} & \multicolumn{1}{p{1.5cm}|}{$N/A$} & 
		\multicolumn{1}{p{1.8cm}||}{$N/A$} &
		\multicolumn{1}{p{2.2cm}|}{$N/A$} & \multicolumn{1}{p{2cm}|}{$N/A$} & 
		\multicolumn{1}{p{2cm}|}{$N/A$} & \multicolumn{1}{p{2cm}|}{$N/A$} \\\hline

		\multicolumn{1}{|p{5cm}||}{GTRBAC \cite{JoshiBLG05}} &
		\multicolumn{1}{p{1.3cm}|}{$P/A$} & \multicolumn{1}{p{1.3cm}|}{$P/A$} & 
		\multicolumn{1}{p{0.7cm}||}{$N/A$} &
		\multicolumn{1}{p{1.5cm}|}{$N/A$} & \multicolumn{1}{p{1.5cm}|}{$N/A$} & 
		\multicolumn{1}{p{1.8cm}||}{$N/A$} &
		\multicolumn{1}{p{2.2cm}|}{$N/A$} & \multicolumn{1}{p{2cm}|}{$N/A$} & 
		\multicolumn{1}{p{2cm}|}{$N/A$} & \multicolumn{1}{p{2cm}|}{$N/A$} \\\hline

		\multicolumn{1}{|p{5cm}||}{GEO-RBAC \cite{BertinoCDP05}} &
		\multicolumn{1}{p{1.3cm}|}{$P/A$} & \multicolumn{1}{p{1.3cm}|}{$N/A$} & 
		\multicolumn{1}{p{0.7cm}||}{$N/A$} &
		\multicolumn{1}{p{1.5cm}|}{$N/A$} & \multicolumn{1}{p{1.5cm}|}{$N/A$} & 
		\multicolumn{1}{p{1.8cm}||}{$N/A$} &
		\multicolumn{1}{p{2.2cm}|}{$N/A$} & \multicolumn{1}{p{2cm}|}{$N/A$} & 
		\multicolumn{1}{p{2cm}|}{$N/A$} & \multicolumn{1}{p{2cm}|}{$N/A$} \\\hline

		\multicolumn{1}{|p{5cm}||}{LRBAC \cite{ZhangHS06}} &
		\multicolumn{1}{p{1.3cm}|}{$P/A$} & \multicolumn{1}{p{1.3cm}|}{$P/A$} & 
		\multicolumn{1}{p{0.7cm}||}{$N/A$} &
		\multicolumn{1}{p{1.5cm}|}{$N/A$} & \multicolumn{1}{p{1.5cm}|}{$N/A$} & 
		\multicolumn{1}{p{1.8cm}||}{$N/A$} &
		\multicolumn{1}{p{2.2cm}|}{$N/A$} & \multicolumn{1}{p{2cm}|}{$N/A$} & 
		\multicolumn{1}{p{2cm}|}{$N/A$} & \multicolumn{1}{p{2cm}|}{$N/A$} \\\hline

		\multicolumn{1}{|p{5cm}||}{LoT-RBAC \cite{ChandranJ05}} &
		\multicolumn{1}{p{1.3cm}|}{$P/A$} & \multicolumn{1}{p{1.3cm}|}{$P/A$} & 
		\multicolumn{1}{p{0.7cm}||}{$N/A$} &
		\multicolumn{1}{p{1.5cm}|}{$N/A$} & \multicolumn{1}{p{1.5cm}|}{$N/A$} & 
		\multicolumn{1}{p{1.8cm}||}{$N/A$} &
		\multicolumn{1}{p{2.2cm}|}{$N/A$} & \multicolumn{1}{p{2cm}|}{$N/A$} & 
		\multicolumn{1}{p{2cm}|}{$N/A$} & \multicolumn{1}{p{2cm}|}{$N/A$} \\\hline

		\multicolumn{1}{|p{5cm}||}{X-GTRBAC \cite{BhattiGBJ05}} &
		\multicolumn{1}{p{1.3cm}|}{$P/A$} & \multicolumn{1}{p{1.3cm}|}{$P/A$} & 
		\multicolumn{1}{p{0.7cm}||}{$N/A$} &
		\multicolumn{1}{p{1.5cm}|}{$N/A$} & \multicolumn{1}{p{1.5cm}|}{$N/A$} & 
		\multicolumn{1}{p{1.8cm}||}{$N/A$} &
		\multicolumn{1}{p{2.2cm}|}{$N/A$} & \multicolumn{1}{p{2cm}|}{$N/A$} & 
		\multicolumn{1}{p{2cm}|}{$N/A$} & \multicolumn{1}{p{2cm}|}{$N/A$} \\\hline

		\multicolumn{1}{|p{5cm}||}{\bf Our CAAC Policy Framework} &
		\multicolumn{1}{p{1.3cm}|}{$YES$} & \multicolumn{1}{p{1.3cm}|}{$YES$} & 
		\multicolumn{1}{p{0.7cm}||}{$YES$} &
		\multicolumn{1}{p{1.5cm}|}{$YES$} & \multicolumn{1}{p{1.5cm}|}{$YES$} & 
		\multicolumn{1}{p{1.8cm}||}{$YES$} &
		\multicolumn{1}{p{2.2cm}|}{$YES$} & \multicolumn{1}{p{2cm}|}{$YES$} & 
		\multicolumn{1}{p{2cm}|}{$YES$} & \multicolumn{1}{p{2cm}|}{$YES$} \\
\hline
\hline
\end{tabular}
\end{center}
\end{table*}

\begin{table*}
\begin{center}
\caption{Comparative analysis of the context-specific RBAC approaches}
\vspace{-0.1in}
\label{table:ca-carbac}
\hspace*{-2.5 in}
\begin{tabular}{|c|c|c|c|c|c|c|c|c|c|c|}
\hline
	\multicolumn{1}{|p{5 cm}||}{\multirow{2}{*} {\bf RBAC Frameworks}} &
		\multicolumn{3}{c||}{\bf Formal Model} &
		\multicolumn{3}{c||}{\bf Implementation Model} &
		\multicolumn{4}{c|}{\bf Evaluation}\\
	\cline{2-11}
	\multicolumn{1}{|p{5cm}||}{\multirow{2}{*} {}} &
		\multicolumn{1}{p{1.3cm}|}{\bf CAURA Policy} & \multicolumn{1}{p{1.3cm}|}{\bf CARPA Policy} & 
		\multicolumn{1}{p{0.7cm}||}{\bf CSL} &
		\multicolumn{1}{p{1.5cm}|}{\bf CAURA Ontology} & \multicolumn{1}{p{1.5cm}|}{\bf CARPA Ontology} & 
		\multicolumn{1}{p{1.8cm}||}{\bf Software Prototype} &
		\multicolumn{1}{p{2.2cm}|}{\bf Completeness} & \multicolumn{1}{p{2cm}|}{\bf Correctness} & 
		\multicolumn{1}{p{2cm}|}{\bf Consistency} & \multicolumn{1}{p{2cm}|}{\bf Performance} \\
\hline
		\multicolumn{1}{|p{5cm}||}{CA-RBAC \cite{KulkarniT08}} &
		\multicolumn{1}{p{1.3cm}|}{$N/A$} & \multicolumn{1}{p{1.3cm}|}{$P/A$} & 
		\multicolumn{1}{p{0.7cm}||}{$N/A$} &
		\multicolumn{1}{p{1.5cm}|}{$N/A$} & \multicolumn{1}{p{1.5cm}|}{$N/A$} & 
		\multicolumn{1}{p{1.8cm}||}{$N/A$} &
		\multicolumn{1}{p{2.2cm}|}{$N/A$} & \multicolumn{1}{p{2cm}|}{$N/A$} & 
		\multicolumn{1}{p{2cm}|}{$N/A$} & \multicolumn{1}{p{2cm}|}{$N/A$} \\\hline

		\multicolumn{1}{|p{5cm}||}{CERBAC \cite{WangLF08}} &
		\multicolumn{1}{p{1.3cm}|}{$N/A$} & \multicolumn{1}{p{1.3cm}|}{$P/A$} & 
		\multicolumn{1}{p{0.7cm}||}{$N/A$} &
		\multicolumn{1}{p{1.5cm}|}{$N/A$} & \multicolumn{1}{p{1.5cm}|}{$N/A$} & 
		\multicolumn{1}{p{1.8cm}||}{$N/A$} &
		\multicolumn{1}{p{2.2cm}|}{$N/A$} & \multicolumn{1}{p{2cm}|}{$N/A$} & 
		\multicolumn{1}{p{2cm}|}{$N/A$} & \multicolumn{1}{p{2cm}|}{$N/A$} \\\hline

		\multicolumn{1}{|p{5cm}||}{DRBAC \cite{ZhengZZT11}} &
		\multicolumn{1}{p{1.3cm}|}{$N/A$} & \multicolumn{1}{p{1.3cm}|}{$N/A$} & 
		\multicolumn{1}{p{0.7cm}||}{$N/A$} &
		\multicolumn{1}{p{1.5cm}|}{$N/A$} & \multicolumn{1}{p{1.5cm}|}{$N/A$} & 
		\multicolumn{1}{p{1.8cm}||}{$N/A$} &
		\multicolumn{1}{p{2.2cm}|}{$N/A$} & \multicolumn{1}{p{2cm}|}{$N/A$} & 
		\multicolumn{1}{p{2cm}|}{$N/A$} & \multicolumn{1}{p{2cm}|}{$N/A$} \\\hline

		\multicolumn{1}{|p{5cm}||}{Access Control for Web Service \cite{H11}} &
		\multicolumn{1}{p{1.3cm}|}{$N/A$} & \multicolumn{1}{p{1.3cm}|}{$N/A$} & 
		\multicolumn{1}{p{0.7cm}||}{$N/A$} &
		\multicolumn{1}{p{1.5cm}|}{$N/A$} & \multicolumn{1}{p{1.5cm}|}{$N/A$} & 
		\multicolumn{1}{p{1.8cm}||}{$N/A$} &
		\multicolumn{1}{p{2.2cm}|}{$N/A$} & \multicolumn{1}{p{2cm}|}{$N/A$} & 
		\multicolumn{1}{p{2cm}|}{$N/A$} & \multicolumn{1}{p{2cm}|}{$N/A$} \\\hline

		\multicolumn{1}{|p{5cm}||}{Context-Aware RBAC \cite{ScheferWenzlS13}} &
		\multicolumn{1}{p{1.3cm}|}{$N/A$} & \multicolumn{1}{p{1.3cm}|}{$P/A$} & 
		\multicolumn{1}{p{0.7cm}||}{$N/A$} &
		\multicolumn{1}{p{1.5cm}|}{$N/A$} & \multicolumn{1}{p{1.5cm}|}{$N/A$} & 
		\multicolumn{1}{p{1.8cm}||}{$N/A$} &
		\multicolumn{1}{p{2.2cm}|}{$N/A$} & \multicolumn{1}{p{2cm}|}{$N/A$} & 
		\multicolumn{1}{p{2cm}|}{$N/A$} & \multicolumn{1}{p{2cm}|}{$N/A$} \\\hline

		\multicolumn{1}{|p{5cm}||}{Context-Aware RBAC \cite{HosseinzadehVRL16}} &
		\multicolumn{1}{p{1.3cm}|}{$N/A$} & \multicolumn{1}{p{1.3cm}|}{$P/A$} & 
		\multicolumn{1}{p{0.7cm}||}{$N/A$} &
		\multicolumn{1}{p{1.5cm}|}{$N/A$} & \multicolumn{1}{p{1.5cm}|}{$N/A$} & 
		\multicolumn{1}{p{1.8cm}||}{$N/A$} &
		\multicolumn{1}{p{2.2cm}|}{$N/A$} & \multicolumn{1}{p{2cm}|}{$N/A$} & 
		\multicolumn{1}{p{2cm}|}{$N/A$} & \multicolumn{1}{p{2cm}|}{$N/A$} \\\hline

		\multicolumn{1}{|p{5cm}||}{Context-Aware RBAC \cite{TrnkaC16}} &
		\multicolumn{1}{p{1.3cm}|}{$N/A$} & \multicolumn{1}{p{1.3cm}|}{$P/A$} & 
		\multicolumn{1}{p{0.7cm}||}{$N/A$} &
		\multicolumn{1}{p{1.5cm}|}{$N/A$} & \multicolumn{1}{p{1.5cm}|}{$N/A$} & 
		\multicolumn{1}{p{1.8cm}||}{$N/A$} &
		\multicolumn{1}{p{2.2cm}|}{$N/A$} & \multicolumn{1}{p{2cm}|}{$N/A$} & 
		\multicolumn{1}{p{2cm}|}{$N/A$} & \multicolumn{1}{p{2cm}|}{$N/A$} \\\hline

		\multicolumn{1}{|p{5cm}||}{\bf Our CAAC Policy Framework} &
		\multicolumn{1}{p{1.3cm}|}{$YES$} & \multicolumn{1}{p{1.3cm}|}{$YES$} & 
		\multicolumn{1}{p{0.7cm}||}{$YES$} &
		\multicolumn{1}{p{1.5cm}|}{$YES$} & \multicolumn{1}{p{1.5cm}|}{$YES$} & 
		\multicolumn{1}{p{1.8cm}||}{$YES$} &
		\multicolumn{1}{p{2.2cm}|}{$YES$} & \multicolumn{1}{p{2cm}|}{$YES$} & 
		\multicolumn{1}{p{2cm}|}{$YES$} & \multicolumn{1}{p{2cm}|}{$YES$} \\
\hline
\hline
\end{tabular}
\end{center}
\end{table*}

\begin{table*}
\begin{center}
\caption{Comparative analysis of our earlier context/situation-aware RBAC approaches}
\vspace{-0.1in}
\label{table:our-caac}
\hspace*{-2.5 in}
\begin{tabular}{|c|c|c|c|c|c|c|c|c|c|c|}
\hline
	\multicolumn{1}{|p{5 cm}||}{\multirow{2}{*} {\bf RBAC Frameworks}} &
		\multicolumn{3}{c||}{\bf Formal Model} &
		\multicolumn{3}{c||}{\bf Implementation Model} &
		\multicolumn{4}{c|}{\bf Evaluation}\\
	\cline{2-11}
	\multicolumn{1}{|p{5cm}||}{\multirow{2}{*} {}} &
		\multicolumn{1}{p{1.3cm}|}{\bf CAURA Policy} & \multicolumn{1}{p{1.3cm}|}{\bf CARPA Policy} & 
		\multicolumn{1}{p{0.7cm}||}{\bf CSL} &
		\multicolumn{1}{p{1.5cm}|}{\bf CAURA Ontology} & \multicolumn{1}{p{1.5cm}|}{\bf CARPA Ontology} & 
		\multicolumn{1}{p{1.8cm}||}{\bf Software Prototype} &
		\multicolumn{1}{p{2.2cm}|}{\bf Completeness} & \multicolumn{1}{p{2cm}|}{\bf Correctness} & 
		\multicolumn{1}{p{2cm}|}{\bf Consistency} & \multicolumn{1}{p{2cm}|}{\bf Performance} \\
\hline
		\multicolumn{1}{|p{5cm}||}{CAAC \cite{KayesOnt13}} &
		\multicolumn{1}{p{1.3cm}|}{$N/A$} & \multicolumn{1}{p{1.3cm}|}{$P/A$} & 
		\multicolumn{1}{p{0.7cm}||}{$N/A$} &
		\multicolumn{1}{p{1.5cm}|}{$N/A$} & \multicolumn{1}{p{1.5cm}|}{$P/A$} & 
		\multicolumn{1}{p{1.8cm}||}{$P/A$} &
		\multicolumn{1}{p{2.2cm}|}{$N/A$} & \multicolumn{1}{p{2cm}|}{$N/A$} & 
		\multicolumn{1}{p{2cm}|}{$N/A$} & \multicolumn{1}{p{2cm}|}{$N/A$} \\\hline

		\multicolumn{1}{|p{5cm}||}{OntCAAC \cite{KayesHC15CJ}} &
		\multicolumn{1}{p{1.3cm}|}{$N/A$} & \multicolumn{1}{p{1.3cm}|}{$P/A$} & 
		\multicolumn{1}{p{0.7cm}||}{$N/A$} &
		\multicolumn{1}{p{1.5cm}|}{$N/A$} & \multicolumn{1}{p{1.5cm}|}{$P/A$} & 
		\multicolumn{1}{p{1.8cm}||}{$P/A$} &
		\multicolumn{1}{p{2.2cm}|}{$P/A$} & \multicolumn{1}{p{2cm}|}{$N/A$} & 
		\multicolumn{1}{p{2cm}|}{$N/A$} & \multicolumn{1}{p{2cm}|}{$P/A$} \\\hline

		\multicolumn{1}{|p{5cm}||}{PO-SAAC \cite{KayesHC14CAISE}} &
		\multicolumn{1}{p{1.3cm}|}{$N/A$} & \multicolumn{1}{p{1.3cm}|}{$P/A$} & 
		\multicolumn{1}{p{0.7cm}||}{$N/A$} &
		\multicolumn{1}{p{1.5cm}|}{$N/A$} & \multicolumn{1}{p{1.5cm}|}{$P/A$} & 
		\multicolumn{1}{p{1.8cm}||}{$P/A$} &
		\multicolumn{1}{p{2.2cm}|}{$P/A$} & \multicolumn{1}{p{2cm}|}{$N/A$} & 
		\multicolumn{1}{p{2cm}|}{$N/A$} & \multicolumn{1}{p{2cm}|}{$P/A$} \\\hline

		\multicolumn{1}{|p{5cm}||}{Ontological SAAC Framework \cite{KayesHC15IS}} &
		\multicolumn{1}{p{1.3cm}|}{$N/A$} & \multicolumn{1}{p{1.3cm}|}{$P/A$} & 
		\multicolumn{1}{p{0.7cm}||}{$N/A$} &
		\multicolumn{1}{p{1.5cm}|}{$N/A$} & \multicolumn{1}{p{1.5cm}|}{$P/A$} & 
		\multicolumn{1}{p{1.8cm}||}{$P/A$} &
		\multicolumn{1}{p{2.2cm}|}{$P/A$} & \multicolumn{1}{p{2cm}|}{$N/A$} & 
		\multicolumn{1}{p{2cm}|}{$N/A$} & \multicolumn{1}{p{2cm}|}{$P/A$} \\\hline

		\multicolumn{1}{|p{5cm}||}{\bf Our CAAC Policy Framework} &
		\multicolumn{1}{p{1.3cm}|}{$YES$} & \multicolumn{1}{p{1.3cm}|}{$YES$} & 
		\multicolumn{1}{p{0.7cm}||}{$YES$} &
		\multicolumn{1}{p{1.5cm}|}{$YES$} & \multicolumn{1}{p{1.5cm}|}{$YES$} & 
		\multicolumn{1}{p{1.8cm}||}{$YES$} &
		\multicolumn{1}{p{2.2cm}|}{$YES$} & \multicolumn{1}{p{2cm}|}{$YES$} & 
		\multicolumn{1}{p{2cm}|}{$YES$} & \multicolumn{1}{p{2cm}|}{$YES$} \\
\hline
\hline
\end{tabular}
\end{center}
\end{table*}

\end{landscape}
   \clearpage
}

Our analysis focuses primarily on the following {\bf key features/aspects} of our policy framework. We provide (i) {\bf a formal policy model} to represent how to mapping between user and role and mapping between role and permission in accordance with the relevant contextual conditions ({\bf CAURA} and {\bf CARPA policy} models). To this end, we introduce a context specification language ({\bf CSL}) to specify the contextual conditions using the relevant simple and complex context information. We also introduce (ii) {\bf an ontology-based policy model} ({\bf CAURA} and {\bf CARPA ontologies}) and {\bf a software prototype} to realize the formal model. We demonstrate (iii) {\bf the evaluation of the framework} by considering the {\bf completeness}, {\bf correctness}, {\bf consistency}, and {\bf performance} of the policy ontology model semantics.

Tables \ref{table:ca-rbac}, \ref{table:ca-carbac} and \ref{table:our-caac} show all the analysis results in which we use ``{\bf YES}" when a feature is available/implemented in the approach, ``{\bf P/A}" when a feature is partially available in the approach, and ``{\bf N/A}" when a feature is not available in the approach.

\subsubsection{Comparative Analysis of the Spatial and Temporal-Aware RBAC Approaches}


Table \ref{table:ca-rbac} shows a comparative analysis of the RBAC approaches which are composed of spatial and/or temporal information \cite{BertinoBF01, JoshiBLG05, BertinoCDP05, ChandranJ05, ZhangHS06, BhattiGBJ05} with respect to our CAAC policy framework.

In general, the existing spatial and temporal access control approaches consider only specific types of context information, without any {\it context-aware access control (CAAC) model}. The access control policies of these approaches are implemented under the role-based access control and they depend on context information composed of location and/or time information. Different from these extended RBAC approaches in which the role-permission assignments are based on specific types of context information, in our policy framework, both the user-role and role-permission assignments are dynamically performed according to the contextual conditions in terms of diverse context information. Towards this end, we introduce a {\it formal policy model} for specifying the two sets of context-aware access control policies ({\it CAURA} and {\it CARPA} policies), including a {\it context specification language (CSL)} for expressing contextual conditions based on the simple and complex context information. Our policy framework also includes an ontology-based policy model ({\it CAURA} and {\it CARPA} ontologies) to realize the formal model. In addition, we {\it evaluate our policy ontology model} and demonstrate its effectiveness.

%
%
%

\subsubsection{Comparative Analysis of the Context-Specific RBAC Approaches}

Table \ref{table:ca-carbac} shows a comparative analysis of the existing context-specific RBAC approaches \cite{KulkarniT08, WangLF08, ZhengZZT11, H11, ScheferWenzlS13, HosseinzadehVRL16, TrnkaC16} with respect to our CAAC policy framework.

Similar to spatial and temporal-aware RBAC approaches, the existing context-specific RBAC approaches support the specification of access control policies in terms of specific types of context information, each of them having different goals and they are highly domain-specific. However, there is still a need for a {\it context-aware access control (CAAC) framework} to incorporate the relevant context information into the basic RBAC processes. To fill this gap, in this paper we have introduced both the {\it formal and ontology-based policy models} to specify {\it context-aware user-role assignments (CAURA)} and {\it context-aware role-permission assignments (CARPA)} policies. Using these models, we provide the {\it architecture and (software) prototype implementation} for building CAAC applications that enforces CAURA and CARPA policies. We present the {\it evaluation results} of the policy ontology model and framework through the prototype and the results demonstrate the {\it completeness}, {\it correctness} and {\it consistency} of the ontology model concepts/semantics, and its {\it efficiency} in terms of response time as well.

\subsubsection{Comparative Analysis of Our Earlier Context/Situation-Aware RBAC Approaches}

Table \ref{table:our-caac} shows a comparative analysis of our earlier work on context/situation-aware access control \cite{KayesOnt13, KayesHC15CJ, KayesHC14CAISE, KayesHC15IS} with respect to our CAAC policy framework.

The CAAC policy framework presented in this paper extends our earlier work on access control in the following ways. In general, our previous context/situation-aware access control approaches support the specification and enforcement of RBAC policies in terms of relevant context information. However, these access control approaches and their associated policy models are still limited in representing and modelling both the two sets of context-aware access control policies (context-aware user-role and role-permission assignments policies). As a consequence, they are not able to offer the advantages of context-aware user-role and role-permission assignments in accordance with the relevant contextual conditions. To this end, we introduce a formal policy model to specify the two sets of policies. The formal model also includes a simple language to express contextual conditions based on the simple and complex context information. In addition, in this paper we extend our previous software architecture and prototype \cite{KayesHC15CJ, KayesHC15IS} for building CAAC applications, providing mechanisms and tool supports for modelling and enforcing CAURA and CAPRA policies. Furthermore, this paper justifies the feasibility of the policy ontology model by demonstrating its completeness, correctness, consistency and efficiency, whereas our earlier access control approaches demonstrate the feasibility in terms of the case study and performance overhead.


%

%



\section{Conclusion and Future Work}
\vspace{-0.1in}
\label{cfw}

A general policy framework is presented in this paper for context-aware access control following the role-based access control mechanism. Our framework significantly differs from the existing access control frameworks in that it considers context-aware user-role and role-permission assignments and consequently supports context-specific access control to resources by leveraging the dynamically changing context information. We have presented a formal model for specifying the context-aware access control policies in our framework. By introducing the concepts of context-aware user-role and role-permission assignments, the association of users to roles can be achieved dynamically based on the relevant context information, and these users can access the resources through the further dynamic association of context-dependent roles to permissions. 

Based on the formal model, we have developed an ontology-based policy framework for modelling and enforcing the two sets of access control policies: the context-aware user-role and role-permission assignment policies. The first set of policies specifies that users can play a particular role when certain contextual conditions are satisfied. The second set of policies specifies that users having roles are allowed to carry out an operation on resources when some further contextual conditions are satisfied. When a user wants to access resources at runtime, policy enforcement determines whether or not an access request is granted or denied. 

We have demonstrated the feasibility of the proposed framework by considering factors like (i) the completeness of the ontology concepts, (ii) the correctness and consistency of the ontology semantics, and (iii) the performance of the framework. The evaluation results show that the core policy ontology with its domain-specific ontologies not only deliver {\it complete}, {\it correct} and {\it consistent} semantics but also demonstrate its {\it efficiency} in terms of response time.



\nocite{*}
\bibliographystyle{compj}
\bibliography{bib}

\vspace {1ex}

\section{Appendix A: CAURA policy specification}

The following code fragment in OWL (see Definition \ref{bcd}) shows the definition of all basic classes of CAURA policy: {\it CAURAPolicy}, {\it ContextualCondition}, {\it ContextInfo}, {\it User}, and {\it Role} (see {\it CAURAPolicy} ontology in Figure \ref{caurao} in Section \ref{caurapo}).

\begin{defn}
(Definitions of Basic Classes).
\label{bcd}
\end{defn} 
{\it \hspace*{0.05 in}$<$owl:Class rdf:ID=``CAURAPolicy"$>$\\
\hspace*{0.2 in}$<$owl:Class rdf:ID=``ContextualCondition"$>$\\
\hspace*{0.2 in}$<$owl:Class rdf:ID=``ContextInfo"$>$\\
\hspace*{0.2 in}$<$owl:Class rdf:ID=``User"$>$\\
\hspace*{0.2 in}$<$owl:Class rdf:ID=``Role"$>$}\\

Definition \ref{huid} shows the class {\it CAURAPolicy} has an object property $hasUser$, which is used to link the classes {\it CAURAPolicy} and {\it User}.

\begin{defn}
(`hasUser' Object Property Definition).
\label{huid}
\end{defn} 
{\it       
\hspace*{0.05 in}$<$owl:ObjectProperty rdf:ID=``hasUser"$>$\\
\hspace*{0.3 in}$<$rdfs:domain rdf:resource=``\#CAURAPolicy"/$>$\\
\hspace*{0.3 in}$<$rdfs:range rdf:resource=``\#User"/$>$\\
\hspace*{0.2 in}$<$/owl:ObjectProperty$>$}\\ 

Similar to Definition \ref{huid}, we define two other object properties $hasRole$ and $hasCondition$. The property $hasRole$ is used to link the classes $CAURAPolicy$ and $Role$ (see Definition \ref{hr}), and the property $hasCondition$ links the classes $CAURAPolicy$ and $ContextualCondition$ (see Definition \ref{hud}).

\begin{defn}
(`hasRole' Object Property Definition).
\label{hr}
\end{defn} 
{\it  \hspace*{0.05 in}$<$owl:ObjectProperty rdf:ID=``hasRole"$>$\\
\hspace*{0.3 in}$<$rdfs:domain rdf:resource=``\#CAURAPolicy"/$>$\\
\hspace*{0.3 in}$<$rdfs:range rdf:resource=``\#Role"/$>$\\
\hspace*{0.2 in}$<$/owl:ObjectProperty$>$}\\

\begin{defn}
(`hasCondition' Object Property Definition).
\label{hud}
\end{defn} 
{\it  \hspace*{0.05 in}$<$owl:ObjectProperty rdf:ID=``hasCondition"$>$\\
\hspace*{0.3 in}$<$rdfs:domain rdf:resource=``\#CAURAPolicy"/$>$\\
\hspace*{0.3 in}$<$rdfs:range rdf:resource=``\#ContextualCondition"/$>$\\
\hspace*{0.2 in}$<$/owl:ObjectProperty$>$}\\

Definition \ref{uid} shows that the class {\it User} has a data type property $userIdentity$, which is {\it xsd:string} type, while  Definition \ref{ridd} shows that the class {\it Role} has a {\it xsd:string} type property, named $roleIdentity$.

\begin{defn}
(`userIdentity' Data Type Property Definition).
\label{uid}
\end{defn}
{\it \hspace*{0.05 in}$<$owl:DatatypeProperty rdf:ID=``userIdentity"$>$\\
\hspace*{0.3 in}$<$rdfs:domain rdf:resource=``\#User"/$>$\\
\hspace*{0.3 in}$<$rdfs:range rdf:resource=``\&xsd;string"/$>$\\
\hspace*{0.2 in}$<$/owl:DatatypeProperty$>$}\\

\begin{defn}
(`roleIdentity' Data Type Property Definition).
\label{ridd}
\end{defn}
{\it \hspace*{0.05 in}$<$owl:DatatypeProperty rdf:ID=``roleIdentity"$>$\\
\hspace*{0.3 in}$<$rdfs:domain rdf:resource=``\#Role"/$>$\\
\hspace*{0.3 in}$<$rdfs:range rdf:resource=``\&xsd;string"/$>$\\
\hspace*{0.2 in}$<$/owl:DatatypeProperty$>$}\\

Definition \ref{hasob} shows the class {\it ContextualCondition} has an object property $hasContext$, which is used to link the classes {\it ContextualCondition} and {\it ContextInfo}.

\begin{defn}
(`hasContext' Object Property Definition).
\label{hasob}
\end{defn} 
{\it  \hspace*{0.05 in}$<$owl:ObjectProperty rdf:ID=``hasContext"$>$\\
\hspace*{0.3 in}$<$rdfs:domain rdf:resource=``\#ContextualCondition"/$>$\\
\hspace*{0.3 in}$<$rdfs:range rdf:resource=``\#ContextInfo"/$>$\\
\hspace*{0.2 in}$<$/owl:ObjectProperty$>$}\\

Definition \ref{ci2s} shows that the class {\it ContextInfo} has two subclasses {\it SimpleContext} and {\it ComplexContext}.

\begin{defn}
(ContextInfo Class and Its Two Subclasses).
\label{ci2s}
\end{defn} 
{\it \hspace*{0.05 in}$<$owl:Class rdf:ID=``SimpleContext"$>$\\
 \hspace*{0.3 in}$<$rdfs:subClassOf rdf:resource=``\#ContextInfo"/$>$\\
\hspace*{0.2 in}$<$/owl:Class$>$\\
 \hspace*{0.2 in}$<$owl:Class rdf:ID=``ComplexContext"$>$\\
  \hspace*{0.3 in}$<$rdfs:subClassOf rdf:resource=``\#ContextInfo"/$>$\\
\hspace*{0.2 in}$<$/owl:Class$>$}\\

\section{Appendix B: CARPA policy specification}

The following code fragment in OWL (see Definition \ref{bcd1}) shows the definition of all basic classes of CARPA policy: {\it CARPAPolicy}, {\it Permission}, {\it Operation}, {\it Resource}, {\it Owner}, and {\it AccessDecision} (see {\it CARPAPolicy} ontology in Figure \ref{f:carpao} in Section \ref{carpapo}). Note that we have already defined the classes {\it Role}, {\it ContextualCondition}, {\it ContextInfo}, {\it SimpleContext}, and {\it ComplexContext} in Appendix A.
\begin{defn}
(Definition of Basic Classes).
\label{bcd1}
\end{defn} 
{\it \hspace*{0.05 in}$<$owl:Class rdf:ID=``CARPAPolicy"$>$\\
\hspace*{0.2 in}$<$owl:Class rdf:ID=``Permission"$>$\\
\hspace*{0.2 in}$<$owl:Class rdf:ID=``Operation"$>$\\
\hspace*{0.2 in}$<$owl:Class rdf:ID=``Resource"$>$\\
\hspace*{0.2 in}$<$owl:Class rdf:ID=``Owner"$>$\\
\hspace*{0.2 in}$<$owl:Class rdf:ID=``AccessDecision"$>$}\\

Each {\it AccessDecision} class instance has exactly one {\it decision} attribute value, which means the value of the {\it decision} attribute may be ``Granted" or ``Denied". As such, it is not possible for the {\it decision} attribute to have both ``Granted" and ``Denied" values for an access policy. Definition \ref{dcd} specifies the cardinality of the class {\it AccessDecision} on the property $decision$.

\begin{defn}
(Cardinality Constraint Definition of a Property `decision').
\label{dcd}
\end{defn}
{\it \hspace*{0.05 in}$<$owl:Class rdf:ID=``AccessDecision"$>$\\
\hspace*{0.3 in}$<$owl:Restriction$>$\\
\hspace*{0.4 in}$<$owl:onProperty rdf:resource=``\#decision"/$>$\\
\hspace*{0.4 in}$<$owl:cardinality rdf:datatype=\\\hspace*{0.4 in}``\&xsd;nonNegativeInteger"$>$1\\
\hspace*{0.4 in}$<$/owl:cardinality$>$\\
\hspace*{0.3 in}$<$/owl:Restriction$>$\\
\hspace*{0.2 in}$<$/owl:Class$>$}\\

The following OWL code shows that the object property $hasDecision$ links the classes $CARPAPolicy$ and $AccessDecision$ (see Definition \ref{hdd}).

\begin{defn}
(`hasDecision' Object Property Definition).
\label{hdd}
\end{defn} 
{\it 
\hspace*{0.05 in}$<$owl:ObjectProperty rdf:ID=``hasDecision"$>$\\
\hspace*{0.3 in}$<$rdfs:domain rdf:resource=``\#CARPAPolicy"/$>$\\
\hspace*{0.3 in}$<$rdfs:range rdf:resource=``\#AccessDecision"/$>$\\
\hspace*{0.2 in}$<$/owl:ObjectProperty$>$}\\

The class $Permission$ links to the classes $Resource$ and $Operation$ using two object properties $hasResource$ and $hasOperation$, respectively (see Definitions \ref{hasres} and \ref{hasop}).

\begin{defn}
(`hasResource' Object Property Definition).
\label{hasres}
\end{defn} 
{\it  
\hspace*{0.05 in}$<$owl:ObjectProperty rdf:ID=``hasResource"$>$\\
\hspace*{0.3 in}$<$rdfs:domain rdf:resource=``\#Permission"/$>$\\
\hspace*{0.3 in}$<$rdfs:range rdf:resource=``\#Resource"/$>$\\
\hspace*{0.2 in}$<$/owl:ObjectProperty$>$}\\

\begin{defn}
(`hasOperation' Object Property Definition).
\label{hasop}
\end{defn} 
{\it 
\hspace*{0.05 in}$<$owl:ObjectProperty rdf:ID=``hasOperation"$>$\\
\hspace*{0.3 in}$<$rdfs:domain rdf:resource=``\#Permission"/$>$\\
\hspace*{0.3 in}$<$rdfs:range rdf:resource=``\#Operation"/$>$\\
\hspace*{0.2 in}$<$/owl:ObjectProperty$>$}\\

The $Operation$ class has a data type property $action$ ({\it xsd:string} type) in order to capture the operation on the resource. The following OWL code shows the property definition (see definition \ref{acdef}). 

\begin{defn}
(`action' Data Type Property Definition).
\label{acdef}
\end{defn}
{\it \hspace*{0.05 in}$<$owl:DatatypeProperty rdf:ID=``action"$>$\\
\hspace*{0.3 in}$<$rdfs:domain rdf:resource=``\#Operation"/$>$\\
\hspace*{0.3 in}$<$rdfs:range rdf:resource=``\&xsd;string"/$>$\\
\hspace*{0.2 in}$<$/owl:DatatypeProperty$>$}\\

The class $Resource$ links to the $Owner$ class using an object property {\it isOwnerBy} (see Definition \ref{isownedby}).

\begin{defn}
(`isOwnedBy' Object Property Definition).
\label{isownedby}
\end{defn} 
{\it \hspace*{0.05 in}$<$owl:ObjectProperty rdf:ID=``isOwnedBy"$>$\\
\hspace*{0.3 in}$<$rdfs:domain rdf:resource=``\#Resource"/$>$\\
\hspace*{0.3 in}$<$rdfs:range rdf:resource=``\#Owner"/$>$\\
\hspace*{0.2 in}$<$/owl:ObjectProperty$>$}\\

The class {\it Resource} has a data type property $resourceIdentity$, which is of {\it xsd:string} type (see Definition \ref{rid}), and the class {\it Owner} has a property $ownerIdentity$ of {\it xsd:string} type (see Definition \ref{oid}).

\begin{defn}
(`resourceIdentity' Data Type Property Definition).
\label{rid}
\end{defn}
{\it \hspace*{0.05 in}$<$owl:DatatypeProperty rdf:ID=``resourceIdentity"$>$\\
\hspace*{0.3 in}$<$rdfs:domain rdf:resource=``\#Resource"/$>$\\
\hspace*{0.3 in}$<$rdfs:range rdf:resource=``\&xsd;string"/$>$\\
\hspace*{0.2 in}$<$/owl:DatatypeProperty$>$}\\

\begin{defn}
(`ownerIdentity' Data Type Property Definition).
\label{oid}
\end{defn}
{\it \hspace*{0.05 in}$<$owl:DatatypeProperty rdf:ID=``ownerIdentity"$>$\\
\hspace*{0.3 in}$<$rdfs:domain rdf:resource=``\#Owner"/$>$\\
\hspace*{0.3 in}$<$rdfs:range rdf:resource=``\&xsd;string"/$>$\\
\hspace*{0.2 in}$<$/owl:DatatypeProperty$>$}\\

\vspace {2in}

\noindent {\bf{A. S. M. Kayes}} is a Postdoctoral Research Fellow in the Department of Computer Science and Information Technology, La Trobe University, Australia. He received his PhD in Computer Science and Software Engineering from Swinburne University of Technology, Australia in 2014, and his BSc in Computer Science and Engineering from Chittagong University of Engineering and Technology, Bangladesh in 2005. His research interests include information modelling, context-aware security, predictive analytics, privacy protection, IoTs and big data integration.
\\
\\
\noindent {\bf{Jun Han}} is a Professor of Software Engineering in the Faculty of Science, Engineering and Technologies at Swinburne University of Technology, Australia. He has also been a research leader with Australia's Cooperative Research Centre in Smart Services (Smart Services CRC) and Cooperative Research Centre in Advanced Automotive Technology (AutoCRC). He received his BEng and MEng in Computer Science and Engineering from Beijing University of Science and Technology in 1982 and 1986 respectively, and his PhD in Computer Science from the University of Queensland in 1992. His research interests include adaptive and context-aware software systems, software and system architectures, software security and performance, services engineering and management, and system integration, evolution and interoperability. He has published more than 220 peer-reviewed papers in international journals and conference proceedings.
\\
\\
\noindent {\bf{Wenny Rahayu}} is a Professor and the Head of School of Engineering and Mathematical Sciences at La Trobe University, Australia. Prior to this appointment, she was the Head of Department of Computer Science and Information Technology from 2012-2014. She received a PhD in Computer Science from La Trobe university in 2000. The main focus of her research is the integration and consolidation of heterogeneous data and systems to support a collaborative environment within a highly data-rich environment. In the last 10 years, she has published two authored books, three edited books and more than 150 research papers in international journals and conference proceedings.
\\
\\
\noindent {\bf{Md Saiful Islam}} is a Lecturer in the School of Information and Communication Technology, Griffith University, Australia. He has finished his PhD in Computer Science and Software Engineering from Swinburne University of Technology, Australia in 2014. He has received his BSc and MS degree in Computer Science and Engineering from University of Dhaka, Bangladesh, in 2005 and 2007, respectively. His current research interests are in the areas of database usability, spatial data management and big data analytics. 
\\
\\
\\
\\
\noindent {\bf{Alan Colman}} is a Senior Lecturer in the Faculty of Science, Engineering and Technologies at Swinburne University of Technology, Australia. He received his MS and PhD degrees from Swinburne University of Technology, Australia. His research interests include adaptive and goal-oriented software architecture, organizational software abstractions, support for user autonomy in complex systems, context aware systems, and performance of prediction and management of service based systems. He has published more than 120 peer-reviewed papers in international journals and conference proceedings.

\end{document}